# Impact of the political risk on food reserve ratio: evidence across countries

Kai Xing[a], Shang Li[b], Xiaoguang Yang[c,d,1]

*a School of Economics & Management, Nanchang University, China*
*b Department of Statistics and Data Science, National University of Singapore, Singapore*
*c Academy of Mathematics and Systems Science, Chinese Academy of Sciences, China*
*d University of Chinese Academy of Sciences, China*

## Abstract

Using an unbalanced panel data covering 75 countries from 1991 to 2019, we explore how the political risk impacts on food reserve ratio. The empirical findings show that an increasing political risk negatively affect food reserve ratio, and same effects hold for both internal risk and external risk. Moreover, we find that the increasing external or internal risks both negatively affect production and exports, but external risk does not significantly impact on imports and it positively impacts on consumption, while internal risk negatively impacts on imports and consumption. The results suggest that most of governments have difficulty to raise subsequent food reserve ratio in face of an increasing political risk, no matter it is an internal risk or an external risk although the mechanisms behind the impacts are different.

* The research is supported by NSFC (72192800)
[1] Corresponding author. Tel.: +86 10 82541810
  *Email addresses*: xgyang@iss.ac.cn (X. G. Yang), xingkai@ncu.edu.cn (K. Xing), lishang@u.nus.edu (S. Li)

# 1. Introduction

In this paper, we investigate whether and how the political risk impacts on food reserve ratio over the period of 1991 to 2019. Our motivation for conducting this research is mainly from a straightforward intuition as follows.

Once the political risk goes upward, the food security will be increasingly threatened. Then the governments will increase food reserve ratio, which is a basic guarantee for food security. This idea reflects a naive logic, which is, people would increase the awareness of risk prevention and raise the abilities of managing the risks after encountering risks.

The intuition could be easily found in the real world. For example, after suffering from natural disasters, people tend to build more solid infrastructure and develop early warning system. The 2006 Yogyakarta earthquake occurred on 27 May with a moment magnitude of 6.4, which killed about 6,000 people and destroyed more than 60,000 houses in the city. After the earthquake, more than 100,000 strengthened houses were constructed using earthquake-resistant technology within two years. Similarly, after the Cyclone BOB06 in 1999, Indian government established the Odisha State Disaster Mitigation Authority (OSDMA) to help preventing losses from cyclone. Corresponding measures include adding new evacuation roads and bridges to connect vulnerable communities, improving existing costal embankments, and investing in advanced early warning systems. Consequently, the Cyclone Phailin in 2013 only killed 50 people, which is less than 1% of Cyclone BOB06 casualties. Analogous to the disasters, people also tend to be more cautious and have quicker risk response capabilities when encountering risks of infectious diseases. Lin et al. (2020) state that in 2003, SARS infected 238 people and killed 33 in Singapore, exposing the weaknesses of response to infectious diseases outbreak. After that, Singapore's healthcare system had drawn to enhance its pandemic preparedness response. Thanks to the previous preparation, Singapore has a case fatality rate of only 0.12% during the COVID-19 pandemic, which is one of the lowest in the world as of 31 Mar 2022.

Meanwhile, the logic of abovementioned reaction-after-risk also holds for political risk. For example, Soviet Union encountered severe famine because of external aggressor. The World War II resulted in massive infrastructural damage in agriculture, and the return of demobilized troops from war and surge in births after the war made the whole system unprepared. These are important reasons for the Soviet famine from 1946 to 1947. Soviet government proposed the Great Plan for the Transformation of Nature to improve agriculture in the nation, which included land development, agricultural practices and water projects. From 1949 to 1965, the Soviet Union planted plantation belts across the entire country along the main rivers and plains in its territory, which solved the problem of insufficient grain production. Although this project is somewhat controversial among the world, the supply of crops, eggs and meat increased from 20% to 100% due to the increase in water supply.

We are interested in examining whether the abovementioned intuition is right or not,

and what happens if it is not true. We investigate how food reserve ratio as a risk management tool for protecting food security, changes along with one year lag of political risk and two types of sub-risks being proxy for internal conflicts and external conflicts[2]. Firstly, we examine whether the governments increase food reserve ratio responding to an increased political risk one year later. Contrary to our naïve intuition, we find the opposite effect. Secondly, we therefore investigate the linkage between political risks and subsequent food reserve ratio in four different scenarios including agriculture machinery level, GDP per capita, food import dependence, and food self-sufficiency, since we are interested in whether and how these scenarios impact on the relationship between political risk and subsequent food reserve ratio. Thirdly, to determine how political risk impacts on subsequent reserve ratio, we further explore the effects of two crucial forces in political risk on subsequent reserve ratio which include external risk (measuring external conflicts) and internal risk (measuring internal conflicts). Finally, we figure out how the abovementioned risks negatively impact on subsequent reserve ratio.

This paper contributes to the literatures in three dimensions. First, to the best of our knowledge, this is the first study showing that the governments could not raise subsequent reserve ratio in face of political risk growth, which contradicts with the naïve intuition. It implies that a government faces difficulty to increase subsequent reserve ratio responding to an increased political risk. This negative effect is robust since this relationship does not change accompanying with change of four economic settings. It is implied that no matter what a country is highly mechanized, developed, less import-dependent, and highly self-sufficient, this country still has difficulty increasing food reserve ratio in face of an increasing political risk. These results correspond to FAO et al. (2019).

Second, there are scarce studies focusing on the comparison between effects of two crucial sub-political risks, i.e. external risk and internal risk, on subsequent reserve ratio (van Weezel, 2018, Koren and Bagozzi, 2016). By providing rigorous empirical evidence, we find that food reserve ratio decreases after appearance of an increased external risk and internal risk. More importantly, we also figure out that the mechanisms about the negative relationships of two risks on subsequent reserve ratio are quite different. We find that both external risk and internal risk have negative relationship with production and exports, indicating governments lose the capacity on the food productions and they should reduce the exports to keep sufficient food on preparing for incoming conflicts. Interestingly, there is quite significant difference between the impact of external and internal risks on domestic consumption and imports, respectively. Internal risk is detrimental to the consumption and imports, while external risk makes consumption increase and does not statistically impact on imports. These results get the support from Ali and Lin (2010).

Third, prior studies have mainly focused on conflict and post-conflict countries, and they investigate how the food reserve and food situation change with the appearance of conflicts or post-conflicts (Koren and Bagozzi, 2016, van Weezel, 2018, Martin-Shields and Stojetz, 2018, Koren, 2018, Jeanty and Hitzhusen, 2006). For example, Jeanty and Hitzhusen (2006) argue that conflicts can initiate food shortages, and disrupt both upstream input markets and downstream output markets, thus deter food production, commercialization and stock management. The conflicts used in their studies are the number of battles related to the death from internal conflicts and external conflicts. However, this method might not give a big picture of internal and external risks. We use the risk index which is able to show the assessment of the political stability for the countries on a comparable basis. This index can cover a complete information for both external and internal conflict risks. Meanwhile, this risk index is calculated by concerning the risks from society, which can better reflect the degree of conflict risk. In this context, we consider the effects of the increasing external risk and internal risk on reserve ratio rather than the effects of incidence of two risks on reserve ratio.

---

[2] These conflicts are represented civil wars, civil disorders, terrorism violence, and wars among countries *etc.*, which are triggered by racialism, gap of wealth, popular sovereignty, unemployment, and foreign pressures *etc.*

The rest of the paper is organized as follows. Next is the literature review, which is followed by the description of the data in section 3. Empirical results and implications are presented in section 4. Section 5 is the robustness tests. The last section is the conclusions and discussions.

## 2. Literature Review

**2.1 Review of the studies about food reserve**

Food reserve has been gradually recognized as a valuable tool for improving accessibility and distribution of food to compensate for offsetting supply shocks or spikes in demand, and facilitating humanitarian response to food emergencies triggered by climate change, water scarcity, natural disasters, and violent conflicts. Food security is better improved nowadays, but we are still in the trouble of facing the threat from food shortage or food insecurity. According to IFAD (2014), there are still 800 million people staying hungry and an extra population burdening. Recently, a group of international organizations, such as the Food and Agriculture Organization (FAO), the World Health Organization (WHO), the United Nations Children Fund (UNICEF) find that there are more than 820 million people being still hungry in the world, which challenges the target of Zero Hungry by 2030 (FAO et al., 2018, FAO et al., 2019). Meanwhile, they also find that the absolute number of undernourished people continually increases. Rooting out hunger is still a globally challengeable issue. 16 October is setted by FAO of UN as World Food Day to raise awareness of food security and hunger (Wang et al., 2020).

The concept of food security has evolved since 1970s. According to Pinstrup-Andersen (2009), its latest concept is determined at World Food Summit (WHS) held in 1996, which is "Food security exists when all people, at all times, have physical and economic access to sufficient safe and nutritious food that meets their dietary needs and food preferences for an active and healthy life." There are four points behind of this definition, which are food availability, food accessibility, food utilization, and food stability.[3] Food reserve is a useful tool to manage the risk of food shortage triggered by problem of food security thereby securing sufficient food for the public. Specifically, the sufficient food reserve plays two major roles in coping with these shocks. One is to be intended to guarantee availability in situations of epidemics, extreme weather or other disasters. Such reserves can help protect the most vulnerable populations; another is to decrease excessive volatility in agricultural commodity markets, since food reserve can not only protect food price crisis from the international market but also help producers to obtain more remunerative prices in such volatility and alleviate the effect of steep price increases on consumers (Lilliston et al., 2012, IATP, 2012).

Wright and Cafiero (2011) investigate the situation of food security in the Middle East and the North Africa (MENA), and they figure out that compared with grain self-sufficiency, a more effective and much cheaper strategy to maintain food security is to ensure food reserves. Caballero-Anthony et al. (2015) reach a similar conclusion, and they argue that there are many approaches to ensure food stability for each country, such as relying on international trade and food markets or improving self-sufficiency level. However, neither has become effective and successful to maintain food security until now. They declare that the preferred approach is to implement a traditional strategy of maintaining food reserves for ensuring greater stability.

Gilbert (2011) investigates whether the trade policy is able to improve food security in developing countries from 2006 to 2011 and finds that maintaining food reserve can be an effective tool to guarantee food security in face of international trade uncertainties and volatility. That is, a country that has sufficient food reserves can be protected from global food price shocks, local supply shocks from poor harvest, income shocks (from economic downturns or exchange rate shocks), disorders in trade due to export bans, as well as during times of emergencies and disasters.

Ancient records show that Chinese emperor has implemented grain reserve policy called "Ever-Normal Granary" since 498 A.D., which was also the oldest model in the world.[4] This policy is intended to maintain food security, which not only guards against famine, but also stabilizes prices for the benefit of both farmers and consumers (Lilliston et al., 2012). Even now, China has consistently kept food reserves proportionately much larger than the rest of the world, which was used as the safeguards to stave off food shortages in order to reduce the negative impact from any unwanted shocks triggered by political risks (Lilliston et al., 2012). Thus, food reserve ratio can be regarded as an important indicator for governments to maintain food security. If food security is threatened, food reserve ratio will increase.

---

[3] Wheeler and Von Braun (2013) detail the four points. Specifically, food availability means "the availability of sufficient quantities of food of appropriate quality, supplied through domestic production or imports"; food accessibility means "access by individuals to adequate resources (entitlements) for acquiring appropriate foods for a nutritious diet"; food utilization means "utilization of food through adequate diet, clean water, sanitation, and health care to reach a state of nutritional well-being where all physiological needs are met"; food stability means "stability, because to be food secure, a population, household or individual must have access to adequate food at all times." Particularly, the last dimension has become a central concern among national governments since stability is considered implicit and necessary for achieving the first three (FAO, 1999).

[4] Keng Shou-chang, as an astronomer and financier in the Western Han Dynasty, develops the food reserve policy called "Ever-Normal Granary" (also named as "constant normal granary"). This policy means that "all the provinces along the boundary of the empire would establish granaries. When the price of grain was low, they should buy it at the normal price, higher than the market price, in order to profit the farmers. When the price was high, they should sell it at the normal price, lower than the market price, in order to profit the consumers", according to Murphy (2009).

**2.2 Review of the studies about political risk**

Political risk covers various components, such as social economic conditions, investment profile, internal and external conflicts. The major important political risks are generated from internal conflict and external conflict. These conflicts related to the issue of controlling territory are the most common political risk (Deaton and Lipka, 2015). In 2005, FAO declared that "conflict and economic problems were cited as the main cause of more than 35 percent of food emergencies between 1992 and 2003, compared to around 15 percent in the period from 1986 to 1991." Meanwhile, under-nourishment is the most predominant conflict in more than half of the countries during the 1990s. FAO also pointed out that civil wars in many developing countries (particularly in sub-Saharan Africa) threaten long-term food security and economic progress (FAO, 2005).

We have seen remarkable upward trend in both the number of conflicts and the severity of war since the end of the Cold War (Dupuy et al., 2017). Numerous studies have identified that these conflicts have detrimental long-run effects on food security represented by nutritional status, consumption, production *etc.* (Martin-Shields and Stojetz, 2018, Breisinger et al., 2015). Dupuy et al. (2017) argue that both internationalized conflicts and non-state conflicts continually threat to global peace. Especially, the pandemic of COVID-19 is nowadays exposing and accelerating trends that have made the world more vulnerable to international conflict (Economists, 2020). Martin-Shields and Stojetz (2019) find that the lives of millions of people around the world are negatively affected by food insecurity, particularly in conflict-affected regions. FAO classifies 19 countries as being in a protracted food crisis which are also impacted by conflict and violence (Holleman et al., 2017). Violent conflict also results in 60 percent of the 815 million undernourished individuals and 79 percent of the 155 million stunted children live in countries (FAO and UNICEF, 2018). It is clear that both food security and economic welfare are threatened by an increasing number of political risks represented by internal and external conflicts. FAO et al. (2019) point out that political instability and conflicts make the problem of hunger deteriorated, resulting in food insecurity. Their study also finds that countries with larger political stability and less violence still have problem of food instability.

We can see an increasing political risk including internal and external conflicts threatens food security since the destruction of agricultural resources or the disruption of food distribution networks and markets (Bora et al., 2010, Messer et al., 2001). For example, Messer et al. (2001) find that an armed conflict results in the destruction of crops, livestock, land, and water, and it also destroys infrastructure, markets, and the human resources required for food production, distribution, and safe consumption. Ali and Lin (2010) find that civil wars significantly push the average cost of food upward. These episodes imply the possible vulnerability of the countries' food system, low agricultural diversity, and the loss of redundancy in the balance between demand and agricultural production (Laio et al., 2016, Timmer, 2010). Messer and Cohen (2015) declare that food and hunger are frequently used as a weapon. The combatants use siege to cut off food supplies and productive capacities, starve opposing populations into submission, and hijack food aid intended for civilians.

## 3. Data

### 3.1 Dataset

We combine country level data from 1991 to 2019 that are made available by a wide variety of sources, including International Country Risk Guide ratings (ICRG) database, FAOSTAT database, Production, Supply, and Distribution (PSD) online database, Emergency Events Database (EM-DAT), and World Bank Open (WBO) database.

ICRG is provided by Political Risk Services (PRS), which gives political, economic, and financial risk ratings and forecasts for its universe of 140 developed, emerging, and frontier markets. Hence, this database allows us to determine the timing and magnitude of changes in a country's political conditions now and in the future. We use a composite political risk and the other two sub-political risks including external risk and internal risk.

PSD online database is constructed by Farm Service Agency (FSA) in United States Department of Agriculture (USDA), and this database provides food data for 176 countries and territories from 1960 to the most recent year available. In particular, PSD database has similar food data as FAOSTAT. Here, we use PSD as this database which covers four years more historical food data in terms of production, stocks, consumption, exports, and imports *etc.* for fifteen types of agricultural commodities in comparison of FAOSTAT. Nine variables are extracted from PSD, which are beginning stocks, ending stocks, domestic consumption, area harvested, yield, imports, exports, Food, Seed and Industrial (FSI) consumption, and production. We use these variables to calculate reserve ratio, import dependency ratio (IDR), and self-sufficiency ratio (SSR).

FAOSTAT database is from FAO, which provides food and agricultural data for over 245 countries and territories, and covers regional groupings from 1961 to the most recent year available. In our study, this database provides three control variables including inland water, producer price index, and number of agricultural machineries.

EM-DAT is launched by the Centre for Research on Epidemiology of Disasters (CRED) starting from 1988. With the initial support of World Health Organization (WHO) and the Belgian Government, this database collects the essential core data on the occurrence and effects of over 22,000 mass disasters from 1900 to the present day. There are various reported disasters, such as drought, flood, extreme temperature, insect infestation, landslide, storm volcanic activity, and wildfires *etc*. We use the total number of disasters, droughts and floods.

WBO database from World Bank (WB) draws from the main WB collections of development indicators, compiled from officially recognized international sources. This database covers global development data, such as World Development Indicators, Health Nutrition and Population Statistics, Worldwide Governance Indicators *etc*. These indicators cover information from over 256 countries and regions, since 1960. We use this database to obtain GDP per capita and population for each country as control variables to allow for the impact from economic level and consumption from the people.

We merge these databases based on the same identifier (country name), and conduct data cleaning. After removing missing values in all variables, we finalize an unbalanced panel dataset including 1922 observations in total, with 75 countries. The list of our sample countries is reported in Table A-2.

**3.2 Variable description**

Basically, we use five staple foods (rice, wheat and three kinds of coarse grains) as the food reserves, following suggestions from Caballero-Anthony et al. (2015). The coarse grains consist of barley, oats, and sorghum. These cereal grains have been widely reserved by the government for the purpose of meeting future domestic or international needs. We then use abovementioned cereal grains to calculate the other variables, such as total value for reserve ratio, yield, area harvested, and producer price index. The dependent variables, key variables and other control variables used in this paper are discussed below. The detailed data description is referred to Table A-1.

3.2.1 Dependent variables

There are five dependent variables including reserve ratio, production, domestic consumption, imports, and exports, respectively. The main dependent variable is the food reserve ratio which is used to proxy for food reserve level. The other four dependent variables are used to support the investigation of mechanisms of how external risk and internal risk impact on food reserve ratio individually. Since these four variables are mainly used to calculate food reserve ratio, we are able to figure out mechanisms behind of both external risk and internal risk affecting reserve ratio via these four variables.[5]

---

[5] Reserve ratio is equal to the beginning stocks divided by domestic consumption. The beginning stocks are closely related with production, imports, exports and consumption. First, the volume of beginning stocks depends on the remainder of grain in the prior year after total supply excluding domestic consumption. Total supply is calculated by production plus imports minus exports. It is clear that if there is a decrease in total supply and an increase in consumption, the less remainder of grains can be allocated for next year's stocks and the beginning of stocks

3.2.2 Key variables

We are interested in finding proxies of political risk, internal risk and external risk with an overwhelming feature of forward looking, which can reflect these risks in a narrow sense. Following Bekaert et al. (2014), the preferred proxies for abovementioned three risks are calculated based on the political risk rating from ICRG, which are political risk index, internal conflict and external conflict, respectively. They are exclusively designed to reflect political risk rather than the other types of country risk, such as economic and financial risk since the ICRG separates these risk individually. Click and Weiner (2010) also suggest that the ICRG rating has capability of differentiating political risk effects. There are plenty of studies using ICRG political risk rating to proxy for various political risks (Caballero et al., 2018, Hasan et al., 2015, Erb et al., 1996, Bekaert et al., 2005).

According to PRS (2021), the political risk rating aims to provide an approach of assessment of the political stability for the countries on a comparable basis, which is done by assigning score to a preset group of factors, termed political risk components. Each component has a minimum score of 0, and maximum score fixed (4, 6 or 12 as introduced below) in the overall political risk assessment. In every case the lower the score total, the higher the risk, and vice versa.

Political risk index is produced by five risk components (including government stability, socioeconomic conditions, investment profile, internal conflict, external conflict, respectively) with score 12, six risk components (including corruption, military in politics, religious tensions, law and order, ethnic tensions, democratic accountability, respectively) with score 6, and one risk component (bureaucracy quality) with weight 4. We use the value of "100 - political risk index" as the proxy for the political risk. Political risk is scored on a scale from 0 (very low risk) to 100 (very high risk).

External conflict is the sub-component of political risk index, which is to assess risk to the incumbent government from foreign action, ranging from non-violent external pressure (diplomatic pressures, withholding of aid, trade restrictions, territorial disputes, sanctions, *etc*.) to violent external pressure (cross-border conflicts to all-out war). The external conflict is further determined by three subcomponents, including war, cross-border conflict, and foreign pressures. We use the value of "12 - external conflict" as the proxy for external risk. External risk is scored on a scale from 0 (very low risk) to 12 (very high risk).

Internal conflict is the sub-component of political risk index, which is to assess political violence in the country and its actual or potential impact on governance, and

---

should be negatively impacted. Second, the measure of reserve ratio should take domestic consumption into consideration. We expect that high domestic consumption reduces the reserve ratio if we fix the beginning of stocks. Combined with two cases, lower beginning of stocks and higher domestic consumption should reduce reserve ratio. Thus, we are able to explore how external and internal risk individually impact on production, imports, exports, and domestic consumption thereby determining their influence on reserve ratio. In fact, ASEAN Food Security Information System (AFSIS) established by the Association of Southeast Asian Nations also monitors and analyzes the abovementioned variables as an early warning mechanism to give the guidance of food reserves level to the alliance member (Lassa et al., 2016).

there are three subcomponents including civil war/coup threat, terrorism/political violence, and civil disorder. The internal conflict is further determined by three subcomponents including civil war/coup threat, terrorism/political violence, and civil disorder. We use the value of "12 - internal conflict" as the proxy for internal risk. Internal risk is scored on a scale from 0 (very low risk) to 12 (very high risk).

3.2.3 Control variables

Prior studies have identified there are many significant factors impacting on food reserve, such as inland water (Finkelshtain et al., 2011, Schneider et al., 2011), population growth (Ehrlich et al., 1993, Godber and Wall, 2014), land sources (harvested area) (Funk and Brown, 2009, Masuda and Goldsmith, 2009), yield (Grassini et al., 2013), price (produced price index) (Lassa et al., 2019, Gilbert, 2011), and natural hazards (Devereux, 2007, Chen and Lan, 2017, Lassa et al., 2019). We use these variables as control variables. [6]

3.2.4 Grouping variables

Inspired by Chen and Lan (2017), we set four economic scenarios to determine whether the effect of political risk on reserve ratio is constant, which are agricultural machinery level, economic development level, food dependence degree, and self-sufficiency degree, respectively. In each scenario, we split the sample into two parts based on the mean value of their proxies[7]. One is the high group; another is the low group. The abovementioned grouping variables are shown as follows.

(1) Agricultural machinery level

Agricultural machinery holds some innovative and effective production technologies, such as breeding, seeding, fertilization, and irrigation. Prior studies identify that agricultural machinery has been regarded as a significantly considerable input on agriculture since it plays a significant role on yield enhancing (Jeanty and Hitzhusen, 2006, Chen and Lan, 2017). It can be seen that a government with utilization of a large number of machineries has a higher grain yield per hectare, leading to the higher food production. This would have positive impact on food reserve ratio.

We select eight kinds of machineries from FAO, which are closely connected with grain production, and sum up to get the total number of machineries. This variable is used to measure agricultural machinery level. If we keep two groups of countries facing the same political risk, we expect the reserve ratio of countries that have more machineries may suffer less than that of counterparties. Intuitively, countries with more machineries have stronger capacity in seeding, irrigating and harvesting of grains, hence they have higher ability to withstand political risk. Hence, countries with higher agricultural machinery level will suffer less from political risk in composition with counterparties.

---

[6] Regarding to the natural hazards, we also try two more disaster types (extreme temperature and insect infestation) that may have direct impact on grains, but both effects on food reserve ratio are not significant.
[7] We also use median value of each proxy to split the sample as a robustness check, and we get similar results.

(2) Economic development level

Jeanty and Hitzhusen (2006) state that economic development and growth is expected to benefit unfortunate countries since strong economy makes these countries have strong purchasing power and advanced technologies. These countries are able to obtain the food in the international market even the food price is high in the international market. Their grain yields increase which enhances labor productivity thereby boosting production. The food reserve ratio can be impacted by economic development. That is, better economic development means more reserve ratio. We follow their approach to use GDP per capita measuring economic development. We can hypothesize the effect of an increased political risk on food reserve ratio would be different between the countries with high GDP per capita and low GDP per capita. People living in countries with high GDP per capita can better resist food insecurity than the counterparties Hence, countries with high GDP per capita would suffer less from rising in political risk than countries with low GDP per capita.

(3) Food dependence degree

This can also be regarded as import dependency, indicating how much of domestic food supply has been imported, or how much domestic food supply comes from the country's own production. If the countries rely on the international market to satisfy their domestic demand, they may suffer a lot under the high volatility of food price. To measure the food dependence degree, we follow Lampietti et al. (2011)'s approach to calculate IDR in Table A-1. Here, the consumption used in this formula is calculated by adding domestic consumption and FSI consumption together using FAO dataset. This indicator provides a measure of the dependence of the countries from food imports. The greater the indicator, the higher the dependence.

We expect that this scenario has strong relationship with reserve ratio to some extents. Specifically, according to Lampietti et al. (2011), the countries with high IDR would be exposure to food price and quantity risk.[8] Timmer (2010) argues that the abovementioned episodes are exposure to systemic vulnerabilities of the global food markets for most net-food importing and lower-income countries. Although these countries can rely on the international markets to secure food safety via trade-based policies, they may find the market closing on them in case of the occurrence of a crisis. Numerous major exporters close their borders by enforcing export bans, which only aggravates the panic and deepens the crisis. We can expect that if countries highly rely on imports, they would suffer more from the exposure to price risk and quantity risk resulted from the political risk erupting.

(4) Self-sufficiency degree

We use SSR to indicate the extent to which a country relies on its own production resources. If a country has ability to meet its domestic food demand to some extents, this country does not necessarily have a high reserve; otherwise, this country should keep sufficient reserve. It is clear that SSR intuitively should be considered as an important scenario that we would test. We follow Michael et al. (2015)'s approach to calculate SSR, which is shown in Table A-1. If a country's SSR increases, this country has a higher self-sufficiency level; otherwise, this country has a lower self-sufficiency level.

---

[8] Price risk means if prices are extremely high, they make purchase difficult even though supplies are available in the world markets. Quantity risk indicates there is no available for the food, even the countries have sufficient funds for purchase.

Intuitively, the countries with higher SSR may not need to lower increase food reserve that much to overcome rise in political risk, since their self-supplied grain may be sufficient to counter the risk so as to weaken this kind of negative impact, while countries with lower SSR may have to draw on their food reserve when political risk increases. We are about to see the countries with the lower SSR suffer more than that with the higher SSR once the political risk happens.

### 3.3 Statistics of variables

Table 7-1 shows summary statistics for each variable. Meanwhile, we use logarithm transformation for all variables to eliminate the effect of different order of magnitude. Considering that zero value may appear, we apply log(1+variable) instead of pure logarithm. Since we are interested in whether the reserve ratio changes with an increased political risk, we allow for a one-year lag in independent variables.

Figure 7-1 Scatter plot for the reserve ratio and political risk/external risk/internal riskshows the scatter plot between the reserve ratio and political risk/external risk/internal risk using sample data in Table A-2. We find that the political risk, external risk, and internal risk are all negatively related with reserve ratio. Figure A-1 World map for three types of risks shows the political risk, external risk, and internal risk around the world using sample data in Table A-2. It can be seen that more developed countries less political risk, external risk, and internal risk.

## 4. Empirical Results and Implications

### 4.1 The relationship between political risk and reserve ratio

Our first question is how food reserve ratio changes along with political risk. We use the following equation to estimate the effect of political risk on reserve ratio.

$$\log(1 + Reserve\ ratio)_{i,t} = \alpha * \log(1 + Political\ risk)_{i,t-1} + \sum_{k=1}^{8} \beta_k * \log(1 + X^k)_{i,t-1} + c_i + year_{t-1} + \varepsilon_{i,t-1} \quad (1)$$

$Political\ risk_{i,t-1}$ is the value of political risk for country $i$ in year $t-1$. $\alpha$, the coefficient of interest, captures the change in $\log(1 + Reserve\ ratio)$ corresponding to the unit change in $\log(1 + Political\ risk)$, indicating how reserve ratio changes along with an increased political risk. A set of year dummies $year_{t-1}$ captures the differences fixed over years. Other control variables that change over time $X^k$, including inland water, area harvested, population, yield, producer price index, disasters, droughts and floods, may affect the volume of food reserve ratio.[9]

Table 7-2 reports the results. In columns (1) and (2), the estimates of $\alpha$ indicate that there is a negative relationship between political risk and reserve ratio, with or without adding control variables. This results is still consistent even using fixed-year effect model in columns (3) and (4). We empirically show that if political risk increases for a country in prior year, food reserve ratio should decrease in current year. This result also gets the support from Figure 7-1. It is suggested that when government faces high political risk, it may have difficulty to raise the level of food reserve. FAO et al. (2019) provide a possible explanation that "political instability limits the capacity of governments to support their populations during food crises, and therefore economic downturns, especially if they are severe, can further compound the impacts of this instability on food crises."

This result also gets the support from the literatures about relationship between political stability and food security. Deaton and Lipka (2015) investigate seven of the most food insecure countries in developing world and use the prevalence of undernourishment from FAO as the proxy for food security. Political stability index is used to represent political stability for each country, which is listed as one of the World Bank's World Wide Governance indicators. They find that political stability is positive related to food security, indicating the rate of undernourishment increases in face of an increasing political risk. This indirectly reflect insufficient food reserve ratio.

Most of the coefficients of control variables correspond to the intuition. Interestingly, droughts do not significantly impact on reserve compared with floods. Devereux (2007) finds that enhancing access to inputs (improved seeds, chemical fertilizers, and tools) for the farmers is an effective approach to boost production and/or reduce crop losses following a weather shock such as droughts. We find that floods are able to positively impact on reserve ratio. That is because department of agriculture and meteorology may predict the floods ahead of time due to the seasonality and regularity of floods, and grains may be reserved in advance for the upcoming floods.

---

[9] We perform the Durbin-Wu-Hausman test, and result shows that fixed effect should be used instead of random effect. In addition, since the difference of political risk among countries are quite small, the results of country fixed effect and the country & year fixed effect are not reported. This is the same case for investigating the impact of internal risk and external risk on reserve ratio.

## 4.2 Further exploration of the linkage between political risk and reserve ratio

We now conclude that high political risk negatively changes along with food reserve ratio. Next, we like to further investigate whether this relationship between political risk and food reserve ratio is unchanged under four economic settings, which are agricultural machinery level (number of machineries), economic development level (GDP per capita), food dependence degree (import dependency ratio), and self-sufficiency degree (self-sufficiency ratio), respectively.

Basically, each part follows the same approach to figure out the effects of political risk on reserve ratio under the corresponding scenario. First, we divide all countries into two opposite groups based on mean value of the proxy of each scenario; Second, we apply Equation (1) in abovementioned two groups; Third, we compare the coefficient of the political risk and its significance level.

### *4.2.1 Agricultural machinery level*

Table 7-3 reports regression results for both more mechanized countries and less mechanized countries. From columns (1) and (2), we can see that in high and low machinery level countries, the coefficients of political risk are both negative. Even when taking year fixed effect into consideration, we have similar results in columns (3) and (4). In addition, political risk has stronger effect on reserve ratio in low machinery level countries, while it has less effect on reserve ratio in high machinery level countries. That is, compared with manual rural activities, the negative effect of increasing political risk is more severe to mechanized agricultural activities, which further lower food reserve ratio. This result corresponds to our intuition. Countries with more machineries have stronger capacity from sowing to harvesting, they have higher ability to fight against political risk.

For control variables, we have the following findings. High mechanization countries have high yield since they have better or advanced agricultural technologies, and their yield has higher positive impact of reserve ratio than low mechanization countries. Compared with low agricultural mechanization countries, change of reserve ratio in high agricultural mechanization countries is less sensitive to the number of area harvested and total disasters. That is because, machines can help to improve efficiency and protect against potential risk (such as emergency harvest right before disasters). Similar reason can be applied to change of population, since only low agricultural mechanization countries need to raise food reserve. For inland water, interestingly, high agricultural mechanization countries can utilize machines to better allocate water resource and further increase grain production, while low agricultural mechanization countries might have water resource allocation problems which may further reduce their food reserve. For the other variables, the results also correspond to our intuition. For example, both two types of countries increase food reserve ratio when producer price index increase.

### *4.2.2 Economic development level*

Table 7-4 reports regression results for both developed and non-developed countries. Columns (1) and (2) show that in both high and low GDP per capita countries, coefficients of political risk are both negative and significant. When we control year fixed effect in columns (3) and (4), our results are still robust. We find that, in non-developed countries, the negative effects of political risk on reserve ratio would be more severe than that in developed countries, which corresponds to our intuition.

For control variables, we have the following findings. Compared with non-developed countries, the change of reserve ratio in developed countries is less sensitive to area harvested, population, yield, producer price index, disasters and floods. In developed countries, people have more capital so that food expenditure only take part of a small proportion of their total expenses. Hence changes of both human and natural factors may not cause huge fluctuations of food reserve ratio. While inland water, as a congenital environment condition, may lead to a different story. For developed countries, more inland water means innately better agricultural production conditions, hence government can raise more reserve; while for non-developed countries, government might not have efficient money to build water conservancy facilities, and the water resource allocation problem may further reduce the reserve ratio.

### *4.2.3 Food dependence degree*

Table 7-5Table 7-5 reports regression results for both high import-dependent countries and less import-dependent countries. Columns (1) and (2) show that both coefficients of political risk are negative, indicating that an increased political risk appears, and the reserve ratio reduces in two types of countries. Columns (3) and (4) show the same results after considering year fixed effect in the model. We can conclude that in both groups of countries, there exists a negative relationship between political risk and reserve ratio. Meanwhile, we find that, in countries which highly rely on imported food, the negative effects of political risk on reserve ratio would be significant and more severe than that in countries with less dependence on imports. This result corresponds to our intuition.

With respective to control variables, we find three variables have different effects on reserve ratio in two groups. In low IDR countries, reserve ratio has significant increase when inland water and area harvested increase, while in high IDR countries, these two local environmental condition factors do not have positive and significant effects, since these countries more relies on external supplies. Also, due to the insufficient production capacity of high IDR countries, government may be more vigilant and raise more reserve when population increase. When facing drought, low IDR countries need to raise food reserve since they need to rely on their own capability to fight against drought, while high IDR countries do not need to do so. The other variables have same effects on food reserve ratio, corresponding our intuition. Interestingly, grain yield in the countries with less IDR is higher than that with more IDR, reflecting countries with less IDR have higher productivity and can acquire the food from their own. When facing disasters and typically floods, high and low IDR countries may take similar actions.

### 4.2.4 *Self-sufficiency degree*

Table 7-6 presents regression results for both countries with high food sufficient level and low food sufficient level. We find that political risk is still significant with negative sign in terms of explaining food reserve ratio in both two groups from columns (1) and (2). Likewise, columns (3) and (4) provide similar conclusion after considering year fixed effect. Particularly, the coefficient of political risk in high SSR countries is smaller than that in low SSR countries, i.e., compared with higher SSR countries, the negative effects of political risk on reserve ratio would be more severe in the countries with lower SSR. This result is consistent with our expectation.

Now we turn to control variables. For inland water, area harvested and population, we find almost the opposite pattern against countries grouping by IDR. This is consistent with our intuition, since countries with high SSR should have less grains imported. We also find that number of droughts have significant positive effect on reserve ratio for high SSR countries, while non-significant effect for low SSR countries. This also does make sense as the low SSR countries may have less drought-resistant technologies and they have nothing to do with droughts. For the remaining variables, high and low SSR countries may have similar responses, corresponding to our intuition.

## 4.3 External risk, internal risk and reserve ratio

This section is to explore two important forces in political risk, i.e., external risk and internal risk, and how they impact on food reserve ratio. We use the following models to investigate the effect of external risk and internal risk on reserve ratio. This model also considers same control variables $X^k$ and a set of year dummies $year_{t-1}$ as Equation (1).

$$\log(1 + Reserve\ ratio)_{i,t} = \alpha * \log(1 + External\ risk)_{i,t-1} + \sum_{k=1}^{8} \beta_k * \log(1 + X^k)_{i,t-1} + c_i + year_{t-1} + \varepsilon_{i,t-1} \quad (2)$$

$$\begin{aligned}
\log(1 + Reserve\ ratio)_{i,t} &= \alpha * \log(1 + Internal\ risk)_{i,t-1} \\
&+ \sum_{k=1}^{8} \beta_k * \log(1 + X^k)_{i,t-1} + c_i + year_{t-1} + \varepsilon_{i,t-1}
\end{aligned} \quad (3)$$

Table 7-7 presents the results. Columns (1) and (7) both show that coefficients of external risk are not significant, no matter consider year fixed effect or not. Columns (2) and (8) both show that coefficients of internal risk is negative and highly significant. Columns (4) and (10) both illustrate negative but not significant coefficients of external risk after adding control variables into the models, no matter consider year fixed effect or not. Columns (5) and (11) illustrate negative and significant coefficients of internal risk after adding control variables into the models. These results are consistent with Figure 7-1 Scatter plot for the reserve ratio and political risk/external risk/internal risk. In columns (3) and (9), both external risk and internal risk are jointly considered for exploring their impacts on reserve ratio. After adding control variables, the behaviors about coefficients of external and internal risk remain unchanged, as shown in columns (6) and (12).

We find that internal risk has much bigger negative impact on reserve ratio than external risk. Ali and Lin (2010) indicate that in the case of internal conflicts, the human suffering is far greater than in the case of external wars since the public faces higher food price or lower purchasing power over food and governments lose the capacity to prioritize and mobilize resources *etc*. This result actually corresponds to Ali and Lin (2010) who state that there is a severe destruction of prevailing economic, social and legal norms in the cases of internal conflicts, while in the event of external conflicts, there is relatively less social and economic destruction. It is worth noting that coefficients of control variables have the expected signs (except inland water and population that have non-significant results) regarding to all of OLS and fixed effect results. Next, we further investigate the mechanism about influence of external risk and internal risk on reserve ratio.

### 4.3.1 *The mechanism of external risk reducing reserve ratio*

We use the following regression to investigate the mechanism of external risk reducing reserve ratio. $Target$ represents four types of dependent variables that we are interested in, which are production, imports, exports and consumption, respectively. This model also considers same control variables $X^k$ and a set of year dummies $year_{t-1}$ as Equation (1).

$$\begin{aligned}
\log(1 + Target)_{i,t} &= \alpha * \log(1 + External\ risk)_{i,t-1} \\
&+ \sum_{k=1}^{8} \beta_k * \log(1 + X^k)_{i,t-1} + c_i + year_{t-1} + \varepsilon_{i,t-1}
\end{aligned} \quad (4)$$

Table 7-8shows the results. A group of models in columns (1), (3), (5) and (7) provide the results of the effects of external risk on production, imports, exports, consumption, respectively. Another group of models in columns (2), (4), (6) and (8) also consider control variables. Simultaneously, after we allow for year fixed effect, our results are robust, see columns (10), (12), (14) and (16). We conclude that external risk negatively impacts on production and exports, and it positively impacts on consumption. Particularly, there is no impact for external risk on imports. Therefore, reserve ratio decreases since production decreases, exports decrease, and consumption increase. The signs of control variables in most columns are in line with our expectation.

### 4.3.2 *The mechanism of internal risk reducing reserve ratio*

We use following regression to investigate the mechanism of internal risk reducing reserve ratio. $Target$ are four types of dependent variables that we are interested in, and they are production, imports, exports, and consumption. This model also considers same control variables $X^k$ and a set of year dummies $year_{t-1}$ as Equation (1).

$$\begin{aligned}
\log(1 + Target)_{i,t} &= \alpha * \log(1 + Internal\ risk)_{i,t-1} \\
&+ \sum_{k=1}^{8} \beta_k * \log(1 + X^k)_{i,t-1} + c_i + year_{t-1} + \varepsilon_{i,t-1}
\end{aligned} \quad (5)$$

Table 7-9 shows the results. A group of models in columns (1), (3), (5) and (7) provide results of the effects of internal risk on production, imports, exports, and consumption, respectively. Another group of models in columns (2), (4), (6) and (8) also consider control variables. Simultaneously, after we allow for year fixed effect, our results are robust, see columns (10), (12), (14) and (16). We can conclude that internal risk negatively impacts on production, imports, exports, and consumption. In order to identify how reserve ratio changes, we further construct following models.

$$\begin{aligned}
\log(1 + Reserve\ ratio)_{i,t} &= \alpha * \log(1 + RSD)_{i,t-1} \\
&+ \sum_{k=1}^{8} \beta_k * \log(1 + X^k)_{i,t-1} + c_i + year_{t-1} + \varepsilon_{i,t-1},
\end{aligned} \quad (6)$$

where *RSD* is short for the ratio of total supply and total demand, defined as $(Production_{i,t} + Imports_{i,t})/(Exports_{i,t} + Consumption_{i,t})$. This model also considers same control variables $X^k$ and a set of year dummies $year_{t-1}$ as Equation (1).

Table 7-10 presents the results. We find that the coefficients of RSD are highly significant and positive (larger than 1) in columns (1) and (2). Columns (3) and (4) provide robust results after considering year fixed effect. We can conclude that the changes of production and imports are more rapid than that of exports and consumption. When there is a decrease in production, imports, exports, and consumption, the effects of internal risk on production and imports are higher than that of internal risk on exports and consumption. Therefore, the reserve ratio reduces along with an increase of internal risk. Additionally, the signs of the control variables are most aligned with our intuition.

### *4.3.3 The differences and similarities of mechanisms between external risk and internal risk reducing reserve ratio*

Interestingly, there are distinctive differences in the results of external risk and internal risk on consumption and imports. We find that the external risk can lead to an increase in domestic consumption, and conversely the internal risk can make domestic consumption downward. Meanwhile, there is statistically no impact of external risk on the imports. In contrast, the internal risk has negative impact on imports, indicating that the higher internal risk is, the lower imports are. Both two findings can be explained by Ali and Lin (2010).

For the different impacts of both external and internal risks on the consumption, the food cost[10] triggered by abovementioned two risks plays a key role. Ali and Lin (2010) declare that the civil wars positively affect the food cost, while international wars apparently do not. They find that once a civil war breaks out, policymakers are lack of the resources to exert control on an increasing food cost. An increase in the food cost could result in higher food prices or lower purchasing power over food, either of which may have damaging influences on social and economic well-being. In addition, rent-seeking behavior remarkably increases, particularly in the food sector. Consequently, the public cannot afford the high food cost thereby consuming less food during the civil wars. The food consumption decreases after an increasing internal risk. In contrast, in the event of an international war, governments have a greater capability to prioritize and mobilize resources, implying that they have the capacity to control rent-seeking behavior in food sectors. Furthermore, there is a stability in the labor market in these countries. Women can join into the labor market for the shortage and curb the rising food cost. Particularly, as opposed to internal wars, the foreign aid can enter the market, which is an effective countervailing tool for reducing the rising food cost[11]. Then, the public can afford the food cost thereby

---

[10] According to Ali and Lin (2010), the food cost is defined as the relative wage paid to produce food versus that paid in other sectors.
[11] Ali and Lin (2010) find that the government has capacity to control the boundaries and transport route. The government has the opportunity to acquire the food from the other countries or organization. Consequently, the

consuming more food during the wars. In another words, the food consumption increases after an increasing external risk.

For the different impacts of both external and internal risks on imports, the capacity of managing the resources for a government is a crucial role. Ali and Lin (2010) find that the governments having no capacity of managing the resources in the civil wars indicating they may lose the control of the areas such as ports and border-crossings. They also may not get the resource from the outside. For example, Keen (1994) points out that both the Sudanese government and opposition forces turn food shortage into a weapon to control territories and populations, and restrict access to food aid via attacking relief convoys. The internal risk increases which leads to the decrease in the imports afterwards. In the case of an international war, as opposed to a civil war, as long as the enemies does not control the boundaries and transport route, the government still has the opportunity to acquire food from the other countries or organization. At least, this country is still able to get the food aid as humanitarian assistance from the humanitarian organizations. Thus, the external risk may not impact on imports.

For the similarities of results, both external and internal risk result in reducing production and exports. Specifically, with the perspective of food production, Jeanty and Hitzhusen (2006), Messer and Cohen (2007) and Messer et al. (2001) find that production may drop remarkably in the conflict-affected region. In another word, both internal and external conflicts are negatively correlated with food production since the governments lose the capacity of producing the food, due to the adverse effects on labor supply, access to land and access to credit and/or directs on capital (for example, theft and destruction) (Martin-Shields and Stojetz, 2018, Segovia, 2017, Martin-Shields and Stojetz, 2019). Messer and Cohen (2007) declare that there are cumulative declines in food production and growth rates of food production in 13 out of 14 African conflict countries from 1970 to 1994. With the perspective of food exports, once both external and internal conflicts break out, governments intuitively reduce the exports to maintain food supply in the war time. There would be more devastating impacts on the exports in the event of internal conflicts compared with in the event of external conflicts. This can be explained that the governments have less ability of distributing and managing the sources within the country, reflecting less capacity of controlling the ports, boundaries and transport route in the event of internal conflicts.

---

food aid as humanitarian assistance from the humanitarian organizations, such as World Food Program run by Unite Nations, can enter the country.

# 5. Robustness Tests

## 5.1 Robustness test for grouping variables

We use cross terms regression instead of dividing samples into two groups in section 4.2. The cross terms are designed as political risk multiplied by the four economic settings (or the dummy variable for economic settings). The regression equation regarding to machinery level is

$$\begin{aligned}
\log(1 + Reserve\ ratio)_{i,t} &= \alpha * \log(1 + Political\ risk)_{i,t-1} + \gamma * \log(1 + Machinery)_{i,t-1} \\
&+ \varphi * \log(1 + Political\ risk)_{i,t-1} * \log(1 + Machinery)_{i,t-1} \\
&+ \sum_{k=1}^{8} \beta_k * \log(1 + X^k)_{i,t-1} + c_i + year_{t-1} + \varepsilon_{i,t-1}.
\end{aligned} \quad (7)$$

We can also use dummy variable for machinery level (1 if higher than average), and formulate another regression as

$$\begin{aligned}
\log(1 + Reserve\ ratio)_{i,t} &= \alpha * \log(1 + Political\ risk)_{i,t-1} + \gamma * \log\left(1 + \mathbb{1}_{Machinery}\right)_{i,t-1} \\
&+ \varphi * \log(1 + Political\ risk)_{i,t-1} * \log\left(1 + \mathbb{1}_{Machinery}\right)_{i,t-1} \\
&+ \sum_{k=1}^{8} \beta_k * \log(1 + X^k)_{i,t-1} + c_i + year_{t-1} + \varepsilon_{i,t-1}.
\end{aligned} \quad (8)$$

In equation (10), $Political\ risk_{i,t-1}$ is the value of political risk for country $i$ in year $t-1$. $\alpha$ captures the change in $\log(1 + Reserve\ ratio)$ corresponding to the unit change in $\log(1 + Political\ risk)$, indicating how reserve ratio changes along with an increased political risk. $\gamma$ captures the change in $\log(1 + \mathbb{1}_{Machinery})$ corresponding to the unit change in $\log(1 + Political\ risk)$, where $\mathbb{1}_{Machinery}$ is the indicator function for machinery level. $\varphi$ captures the change in the cross term $\log(1 + Political\ risk) * \log(1 + \mathbb{1}_{Machinery})$ corresponding to the unit change in $\log(1 + Political\ risk)$. A set of year dummies $year_{t-1}$ captures the differences fixed over years. Other control variables that change over time, $X^k$, include inland water, area harvested, population, yield, producer price index, disasters, droughts, and floods, which may affect the volume of food reserve ratio. The empirical results from Table B-1 to Table B-4 convince us that under these four economic settings, political risk still negatively impacts on reserve ratio, indicating that our results are robust.

## 5.2 Test of endogeneity

Despite the significant impact of political risk, external risk and internal risk on food reserve ratio, there might be concerns of the possible endogeneity problem. That is, the reserve ratio may adversely impact on the political risk. For the external conflicts, the countries may fight each other in terms of completing for food supplies. For the internal conflicts, the public may fight against governments as the food is unevenly distributed in the food shortage period. We use the following two methods to handle the abovementioned problem.

### 5.2.1 Lag three years for independent variables

We use same equations as Equation (1) and lag 3 years for all the independent variables. The regression results can be seen in Table C-1 and Table C-2. The coefficients of political risk, external risk and internal risk are still negative and significant. Internal risk has much bigger negative impact on reserve ratio than external risk. Also, most coefficients of control variables are consistent with our intuition.

### 5.2.2 Heckman two-stage procedure

To test the endogeneity that might arise from reverse causality, we implement the Heckman two-stage procedure(Heckman, 1979). The results are from In the first stage, we use the Probit regression model to estimate binary political risk dummy, which equals 1 if political risk is larger than mean value, and 0 otherwise. We add the same control variables as in previous regression, control year fixed effect and generate Inverse Mills ratio (IMR). See Equation (1) for this Probit regression equation. In the second stage, we use the Logit regression model to estimate binary reserve ratio dummy, which equals 1 if reserve ratio is larger than mean value, and 0 otherwise. We add IMR as an independent variable, add same control variables, and also control year fixed effect. Note that in second stage, we allow for one year lag in all independent variables. See Equation (2) for this Logit regression equation. We also replace political risk with external/internal risk and do the same test. The results can be seen in Table C-3, Table C-4 and Table C-5.

$$Probit(binary - Political\ risk)_{i,t} = \sum_{k=1}^{8} \beta_k * \log(1 + X^k)_{i,t} + c_i + year_t + \varepsilon_{i,t} \qquad (1)$$

$$Logit(binary - Reserve\ ratio)_{i,t} = \log(1 + Political\ risk)_{i,t-1} + IMR_{i,t-1} + \sum_{k=1}^{8} \beta_k * \log(1 + X^k)_{i,t-1} + c_i + year_{t-1} + \varepsilon_{i,t-1} \qquad (2)$$

Table C-3 to Table C-5. The coefficients for political risk, external risk and internal risk in second stage are all negative, and coefficient for internal risk is highly significant while not significant for external risk. Particularly, IMR is not significant in all three regressions in second stage, indicating that endogeneity problem does not exist, and the causal relationships from political risk, external and internal risk to reserve ratio are robust.

## 6. Conclusions and Discussions

Using an unbalanced panel data of 75 countries from 1991 to 2019, we investigate the effects of political risk and two crucial sub-political risks (external conflict risk and internal conflict risk) on reserve ratio. The major findings are shown below.

First, we find that reserve ratio decreases when there is an increase in political risk, indicating that governments have no capacity of raising reserve ratio in the context of an increasing political risk. This relationship is robust after considering four scenarios including the number of agricultural machineries, GDP per capita, IDR, and SSR, respectively. Moreover, countries with less agricultural machineries, lower GDP per capita, higher import-dependent ratio suffer more when facing with an increase in political risk, while higher food self-sufficiency countries suffer less.

Second, we find that both external and internal risks negatively impact on reserve ratio, and internal risk has a bigger impact than external risk, implying that countries suffer more in face of an increase in internal risk than external risk. This result corresponds to Ali and Lin (2010)'s findings. Also, we find that there is a significant difference between mechanisms of both impacts of external and internal risk on reserve ratio. External and internal risk are able to reduce production and exports, which is consistent with prior studies. Interestingly, external risk does not statistically impact on imports, and it is positively correlated with consumption. In contrast, internal risk negatively impacts on imports and consumption.

From a policy perspective, empirical results show that once a country's internal risk increases, both the consumption and imports decrease afterwards. It implies that this country is in a big trouble. The food price (purchasing power over food) is higher (lower) since the government has no capacity of managing the resources in the civil wars and the rent-seeking behavior in the food sector remarkably increases. Consumers cannot afford high food cost and consume less food. Besides that, the government's administrative power is impaired, which weakens the capacity to prioritize and mobilize resources. Consequently, the imports decrease with less food supply. The reduction of both food consumption and imports jointly threatens social and economic well-being.

By contrast, once a country's external risk increases, the consumption increases, while imports are uncertain. The food price (purchasing power over food) is lower (higher), since the government has strong ability of organization and greater capability to prioritize and mobilize resources. The consumers can afford the low food cost and consume more food. Meanwhile, government may even obtain food aid as humanitarian assistance from the humanitarian organizations. It is clear that a country facing internal risk suffers more than a country facing external risk, implying that internal stability is more important than external stability for a country. More importantly, the country having internal conflicts should have more urgent and sustainable assistance than that having external conflicts. The international aid organizations, such as United Nations, Red Cross Society, FAO *etc.*, should pay more attention on countries which are in face of internal risk in comparison with external risk.

Our results also have immediate implications for world food security under the pandemic of COVID-19. The epidemic is a serious risk faced by the countries around the world, which affects domestic production, circulation and international supply chain. Boyacı-Gündüz et al. (2021) argue that the impact of COVID-19 pandemic on food systems are mainly from four sectors: production, processing, retailing and consumption. The instability can result in occasional price spikes, market and supply disruptions and food shortages (FAO, 2020). Our research indicates that the food reserve ratio might decrease after a big impact. Therefore, all countries need to make extra efforts to safeguard the food security, e.g., taking special measures to promote domestic production, stopping blaming each other but strengthening the cooperation on international food market, granting the United Nations and other international food or agriculture organizations to play more important roles in food security.

**Word count: 9294 (excluding footnotes); 9994 (including footnotes); They do not include tables/ figures/ maps/ appendices/ references/ abstract/title page/endnotes/acknowledgements.**

Table 7-1 Summary statistics

| Variable | N | Mean | St. Dev. | Min | Median | Max |
|---|---|---|---|---|---|---|
| **Panel A: *Dependent variables*** | | | | | | |
| Reserve ratio | 1,922 | 18.7 | 17.8 | 0 | 14.9 | 149 |
| Production | 1,922 | 14,990.4 | 39,608.8 | 6 | 1,747.5 | 287,214 |
| Total consumption | 1,922 | 20,927.9 | 54,655.8 | 20 | 3,791 | 399,950 |
| Imports | 1,922 | 1,791.0 | 2,714.7 | 0 | 629 | 26,809 |
| Exports | 1,922 | 2,285.5 | 6,573.3 | 0 | 27.5 | 48,227 |
| **Panel B: *Key variables*** | | | | | | |
| Political risk | 1,922 | 36.2 | 11.7 | 7 | 37 | 81 |
| External risk | 1,922 | 2.1 | 1.6 | 0 | 2 | 10 |
| Internal risk | 1,922 | 3.1 | 2.0 | 0 | 3 | 12 |
| **Panel C: *Control variables*** | | | | | | |
| Inland water | 1,922 | 4,798.5 | 15,005.5 | 1 | 600 | 91,416 |
| Area harvested | 1,922 | 5,744.3 | 13,133.9 | 8 | 737.5 | 83,729 |
| Population | 1,922 | 75,214,688.0 | 210,468,402.0 | 783,121 | 15,668,879 | 1,397,715,000 |
| Yield | 1,922 | 8.4 | 5.7 | 0.1 | 7.2 | 25.4 |
| Producer price index | 1,922 | 561.0 | 11,772.0 | 0.0 | 196.4 | 493,743.9 |
| Disasters | 1,922 | 3.5 | 5.4 | 0 | 2 | 43 |
| Droughts | 1,922 | 0.1 | 0.3 | 0 | 0 | 3 |
| Floods | 1,922 | 1.3 | 2.0 | 0 | 1 | 20 |

Figure 7-1 Scatter plot for the reserve ratio and political risk/external risk/internal risk

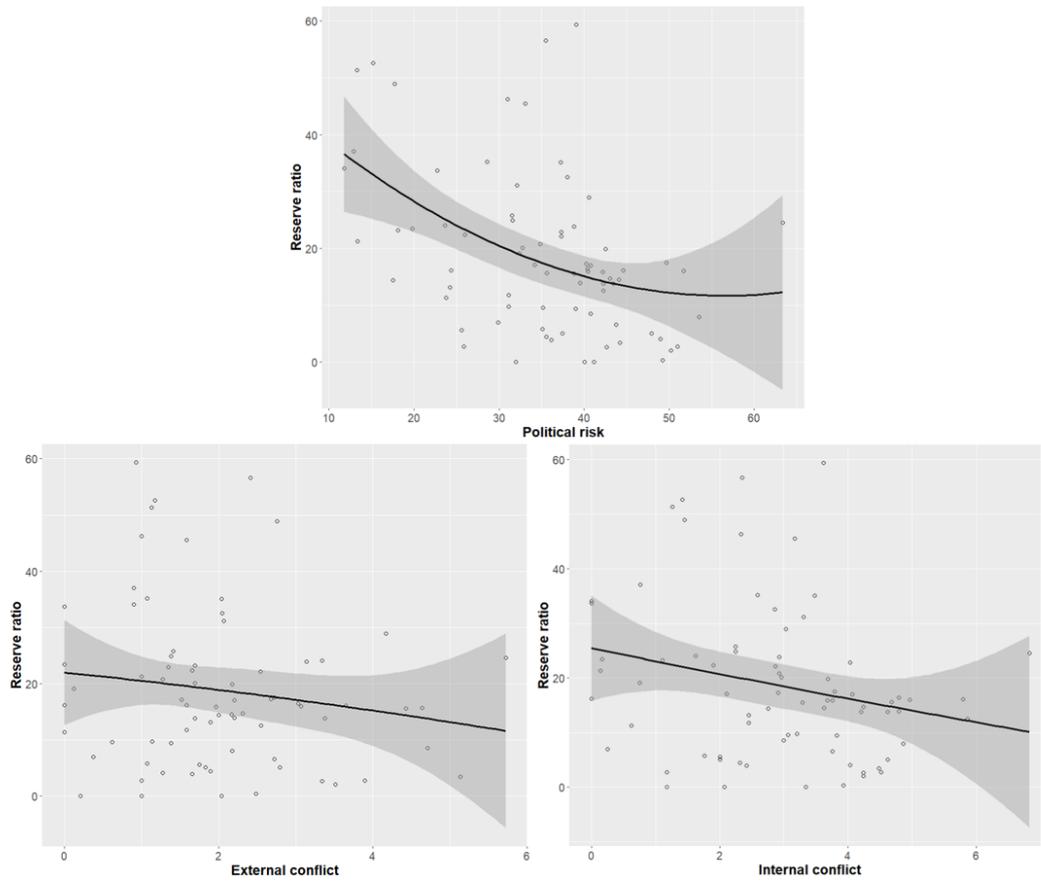

Table 7-2 The effects of political risk on reserve ratio

| | Reserve ratio | | | |
|---|---|---|---|---|
| | (1) | (2) | (3) | (4) |
| Political risk | -1.050*** | -0.572*** | -1.092*** | -0.612*** |
| | (0.072) | (0.072) | (0.072) | (0.074) |
| Inland water | | -0.028* | | -0.030** |
| | | (0.015) | | (0.015) |
| Area harvested | | 0.102*** | | 0.101*** |
| | | (0.021) | | (0.021) |
| Population | | 0.058* | | 0.065* |
| | | (0.034) | | (0.035) |
| Yield | | 0.716*** | | 0.703*** |
| | | (0.044) | | (0.045) |
| Producer price index | | 0.143*** | | 0.147*** |
| | | (0.016) | | (0.019) |
| Disasters | | -0.247*** | | -0.261*** |
| | | (0.052) | | (0.053) |
| Droughts | | 0.194* | | 0.168 |
| | | (0.102) | | (0.103) |
| Floods | | 0.257*** | | 0.276*** |
| | | (0.060) | | (0.061) |
| Year FE | N | N | Y | Y |
| Observations | 1,847 | 1,847 | 1,847 | 1,847 |
| $R^2$ | 0.103 | 0.394 | 0.112 | 0.384 |

Note: In columns (1) and (2), we show the impact of political risk on reserve ratio, with and without adding control variables using OLS regression. In columns (3) and (4), year fixed effect is taken into consideration. The Standard Deviation is reported in round brackets. *, **, and *** are used to denote significance at the 10%, 5%, and 1% level, respectively.

Table 7-3 The effects of political risk on reserve ratio: grouping by agricultural machinery level

| | Reserve ratio | | | |
|---|---|---|---|---|
| Agricultural machinery level | >=mean | <mean | >=mean | <mean |
| | (1) | (2) | (3) | (4) |
| Political risk | -0.203 | -0.654*** | -0.652** | -0.700*** |
| | (0.261) | (0.081) | (0.299) | (0.083) |
| Inland water | 0.255** | -0.052*** | 0.270** | -0.055*** |
| | (0.113) | (0.015) | (0.134) | (0.015) |
| Area harvested | 0.087 | 0.115*** | 0.155 | 0.112*** |
| | (0.153) | (0.022) | (0.168) | (0.022) |
| Population | -0.374* | 0.099*** | -0.465** | 0.111*** |
| | (0.190) | (0.036) | (0.220) | (0.037) |
| Yield | 1.286*** | 0.664*** | 1.014*** | 0.650*** |
| | (0.262) | (0.045) | (0.275) | (0.046) |
| Producer price index | 0.179*** | 0.157*** | 0.194*** | 0.167*** |
| | (0.068) | (0.017) | (0.069) | (0.020) |
| Disasters | -0.141 | -0.220*** | -0.059 | -0.237*** |
| | (0.118) | (0.055) | (0.125) | (0.056) |
| Droughts | -0.233 | 0.149 | -0.357 | 0.117 |
| | (0.346) | (0.106) | (0.373) | (0.107) |
| Floods | 0.128 | 0.191*** | 0.317* | 0.210*** |
|  | (0.148) | (0.064) | (0.173) | (0.065) |
| Year FE | N | N | Y | Y |
| Observations | 129 | 1,718 | 129 | 1,718 |
| R² | 0.796 | 0.395 | 0.846 | 0.386 |

Note: This table shows the result of dividing countries according to agricultural machinery level. This level is defined as the following steps: First, we average the yearly machinery number for each country as a machinery level indicator from 1991 to 2019; Then, we set the mean value of this indicator as a benchmark; Finally, if the number of machineries in a country is higher than benchmark, we define this country as high machinery level country, otherwise it is defined as low machinery level country. Columns (1) and (2) present the result for high and low machinery level countries respectively using OLS regression. In columns (3) and (4), year fixed effect is taken into consideration. The Standard Deviation is reported in round brackets. *, **, and *** are used to denote significance at the 10%, 5%, and 1% level, respectively.

Table 7-4 The effects of political risk on reserve ratio: grouping by GDP per capita

|  | Reserve ratio | | | |
|---|---|---|---|---|
| GDP per capita | >=mean | <mean | >=mean | <mean |
|  | (1) | (2) | (3) | (4) |
| Political risk | -0.511*** | -0.780*** | -0.532*** | -0.892*** |
|  | (0.104) | (0.121) | (0.114) | (0.128) |
| Inland water | 0.127*** | -0.065*** | 0.119*** | -0.067*** |
|  | (0.022) | (0.017) | (0.024) | (0.017) |
| Area harvested | 0.064** | 0.097*** | 0.074** | 0.097*** |
|  | (0.027) | (0.025) | (0.030) | (0.026) |
| Population | 0.046 | 0.074* | 0.043 | 0.082* |
|  | (0.038) | (0.042) | (0.042) | (0.043) |
| Yield | 0.035 | 0.797*** | 0.016 | 0.786*** |
|  | (0.075) | (0.050) | (0.080) | (0.050) |
| Producer price index | 0.078** | 0.149*** | 0.068* | 0.162*** |
|  | (0.035) | (0.018) | (0.038) | (0.021) |
| Disasters | -0.150** | -0.239*** | -0.151** | -0.247*** |
|  | (0.059) | (0.063) | (0.064) | (0.065) |
| Droughts | 0.081 | 0.174 | 0.043 | 0.145 |
|  | (0.141) | (0.116) | (0.154) | (0.118) |
| Floods | -0.001 | 0.273*** | 0.014 | 0.287*** |
|  | (0.067) | (0.073) | (0.076) | (0.075) |
| Year FE | N | N | Y | Y |
| Observations | 320 | 1,527 | 320 | 1,527 |
| R² | 0.536 | 0.353 | 0.527 | 0.341 |

Note: This table shows the result of dividing countries according to GDP per capita level. This level is defined as the following steps: First, average the yearly GDP per capita for each country as a GDP per capita level indicator from 1991 to 2019; Then, set the mean value of this indicator as a benchmark; Finally, if higher than benchmark, define as high GDP per capita level countries, otherwise define as low GDP per capita level countries. Columns (1) and (2) present the result for high and low GDP per capita level countries respectively using OLS regression. In columns (3) and (4), year fixed effect is taken into consideration. There are only global time-series data for Droughts and Floods variables, so no result is presented when controlling for year effect in columns (3) and (4). The Standard Deviation is reported in round brackets. *, **, and *** are used to denote significance at the 10%, 5%, and 1% level, respectively.

Table 7-5 The effects of political risk on reserve ratio: grouping by IDR

|  | Reserve ratio | | | |
|---|---|---|---|---|
| IDR | >=mean | <mean | >=mean | <mean |
|  | (1) | (2) | (3) | (4) |
| Political risk | -0.812*** | -0.074 | -0.846*** | -0.131 |
|  | (0.103) | (0.096) | (0.105) | (0.101) |
| Inland water | -0.085*** | 0.054*** | -0.089*** | 0.048** |
|  | (0.022) | (0.019) | (0.022) | (0.019) |
| Area harvested | 0.001 | 0.267*** | -0.004 | 0.260*** |
|  | (0.044) | (0.031) | (0.045) | (0.031) |
| Population | 0.411*** | -0.232*** | 0.426*** | -0.210*** |
|  | (0.061) | (0.045) | (0.061) | (0.047) |
| Yield | 0.182*** | 1.227*** | 0.158** | 1.214*** |
|  | (0.068) | (0.055) | (0.069) | (0.056) |
| Producer price index | 0.155*** | 0.127*** | 0.179*** | 0.127*** |
|  | (0.025) | (0.020) | (0.028) | (0.024) |
| Disasters | -0.263*** | -0.252*** | -0.290*** | -0.277*** |
|  | (0.076) | (0.064) | (0.077) | (0.067) |
| Droughts | 0.267 | 0.231** | 0.272 | 0.204* |
|  | (0.167) | (0.116) | (0.171) | (0.119) |
| Floods | 0.235** | 0.292*** | 0.286*** | 0.307*** |
|  | (0.092) | (0.072) | (0.096) | (0.075) |
| Year FE | N | N | Y | Y |
| Observations | 942 | 905 | 942 | 905 |
| $R^2$ | 0.329 | 0.562 | 0.330 | 0.551 |

Note: This table shows the result of dividing countries according to IDR. This level is defined as the following steps: First, average the yearly Imports and Consumption for each country from 1991 to 2019; Second, calculate IDR (use the first formula) for each country as a level indicator; Then, set the mean value of this indicator as a benchmark; Finally, if higher than benchmark, define as high IDR level countries, otherwise define as low IDR level countries. Columns (1) and (2) present the result for high and low IDR level countries respectively using OLS regression. In columns (3) and (4), year fixed effect is taken into consideration. The Standard Deviation is reported in round brackets. *, **, and *** are used to denote significance at the 10%, 5%, and 1% level, respectively.

Table 7-6 The effects of political risk on reserve ratio: grouping by SSR

|  | Reserve ratio | | | |
|---|---|---|---|---|
| SSR | >=mean | <mean | >=mean | <mean |
|  | (1) | (2) | (3) | (4) |
| Political risk | -0.269** | -0.781*** | -0.302** | -0.827*** |
|  | (0.130) | (0.086) | (0.142) | (0.088) |
| Inland water | 0.015 | -0.044** | 0.005 | -0.046*** |
|  | (0.029) | (0.017) | (0.029) | (0.018) |
| Area harvested | 0.369*** | 0.028 | 0.350*** | 0.021 |
|  | (0.064) | (0.027) | (0.067) | (0.027) |
| Population | -0.418*** | 0.308*** | -0.377*** | 0.324*** |
|  | (0.071) | (0.044) | (0.076) | (0.045) |
| Yield | 1.054*** | 0.461*** | 1.083*** | 0.438*** |
|  | (0.076) | (0.055) | (0.079) | (0.056) |
| Producer price index | 0.195*** | 0.136*** | 0.232*** | 0.153*** |

|  | (1) | (2) | (3) | (4) |
|---|---|---|---|---|
|  | (0.030) | (0.019) | (0.039) | (0.022) |
| Disasters | -0.221** | -0.216*** | -0.276*** | -0.235*** |
|  | (0.087) | (0.062) | (0.092) | (0.063) |
| Droughts | 0.523*** | 0.105 | 0.477*** | 0.097 |
|  | (0.138) | (0.140) | (0.143) | (0.142) |
| Floods | 0.368*** | 0.209*** | 0.396*** | 0.257*** |
|  | (0.094) | (0.075) | (0.098) | (0.077) |
| Year FE | N | N | Y | Y |
| Observations | 549 | 1,298 | 549 | 1,298 |
| $R^2$ | 0.486 | 0.385 | 0.471 | 0.384 |

Note: This table shows the result of dividing countries according to SSR. This level is defined as the following steps: First, average the yearly SSR for each country as a SSR level indicator from 1991 to 2019; Then, set the mean value of this indicator as a benchmark; Finally, if higher than benchmark, define as high SSR level countries, otherwise define as low SSR level countries. Columns (1) and (2) present the result for high and low SSR level countries respectively using OLS regression. In columns (3) and (4), year fixed effect is taken into consideration. The Standard Deviation is reported in round brackets. *, **, and *** are used to denote significance at the 10%, 5%, and 1% level, respectively.

Table 7-7 The effects of external risk and internal risk on reserve ratio

| | Reserve ratio | | | | | | | | | | | |
|---|---|---|---|---|---|---|---|---|---|---|---|---|
| | (1) | (2) | (3) | (4) | (5) | (6) | (7) | (8) | (9) | (10) | (11) | (12) |
| External risk | 0.043 | | 0.186*** | -0.049 | | -0.014 | -0.051 | | 0.094 | -0.060 | | -0.020 |
| | (0.051) | | (0.057) | (0.042) | | (0.047) | (0.054) | | (0.059) | (0.045) | | (0.048) |
| Internal risk | | -0.213*** | -0.301*** | | -0.090** | -0.084* | | -0.306*** | -0.344*** | | -0.120** | -0.112** |
| | | (0.050) | (0.057) | | (0.044) | (0.048) | | (0.053) | (0.058) | | (0.047) | (0.050) |
| Inland water | | | | 0.001 | -0.002 | -0.002 | | | | 0.0002 | -0.003 | -0.004 |
| | | | | (0.015) | (0.015) | (0.015) | | | | (0.015) | (0.015) | (0.015) |
| Area harvested | | | | 0.102*** | 0.100*** | 0.100*** | | | | 0.100*** | 0.098*** | 0.098*** |
| | | | | (0.021) | (0.021) | (0.021) | | | | (0.021) | (0.021) | (0.021) |
| Population | | | | -0.027 | -0.015 | -0.014 | | | | -0.022 | -0.006 | -0.004 |
| | | | | (0.033) | (0.034) | (0.034) | | | | (0.034) | (0.035) | (0.035) |
| Yield | | | | 0.853*** | 0.834*** | 0.835*** | | | | 0.850*** | 0.824*** | 0.825*** |
| | | | | (0.041) | (0.042) | (0.043) | | | | (0.042) | (0.043) | (0.043) |
| Producer price index | | | | 0.162*** | 0.162*** | 0.162*** | | | | 0.171*** | 0.170*** | 0.170*** |
| | | | | (0.017) | (0.017) | (0.017) | | | | (0.019) | (0.019) | (0.019) |
| Disasters | | | | -0.177*** | -0.183*** | -0.183*** | | | | -0.191*** | -0.199*** | -0.200*** |
| | | | | (0.052) | (0.052) | (0.052) | | | | (0.053) | (0.053) | (0.053) |
| Droughts | | | | 0.133 | 0.135 | 0.135 | | | | 0.107 | 0.107 | 0.107 |
| | | | | (0.104) | (0.104) | (0.104) | | | | (0.105) | (0.105) | (0.105) |
| Floods | | | | 0.220*** | 0.226*** | 0.226*** | | | | 0.238*** | 0.247*** | 0.246*** |
| | | | | (0.061) | (0.061) | (0.061) | | | | (0.062) | (0.062) | (0.062) |
| Year FE | N | N | N | N | N | N | Y | Y | Y | Y | Y | Y |
| Observations | 1,847 | 1,847 | 1,847 | 1,847 | 1,847 | 1,847 | 1,847 | 1,847 | 1,847 | 1,847 | 1,847 | 1,847 |
| $R^2$ | 0.0004 | 0.010 | 0.015 | 0.373 | 0.374 | 0.374 | 0.0005 | 0.018 | 0.020 | 0.362 | 0.363 | 0.363 |

Note: This table presents the results of the effects of external risk and internal risk on reserve ratio. In columns (1) and (4), we show the impact of external risk on reserve ratio, without and with adding control variables using OLS regression. In columns (7) and (10), year fixed effect is taken into consideration. In columns (2) and (5), we show the impact of internal risk on reserve ratio, without and with adding control variables using OLS regression. In columns (8) and (11), year fixed effect is taken into consideration. In columns (3) and (6), we show the impact of external risk and internal risk on reserve ratio, without and with adding control variables using OLS regression. In columns (9) and (12), year fixed effect is taken into consideration. The Standard Deviation is reported in round brackets. ∗, ∗∗, and ∗∗∗ are used to denote significance at the 10%, 5%, and 1% level, respectively.

Table 7-8 The effects of external risk on production, imports, exports, and consumption

| | Production | | Imports | | Exports | | Consumption | | Production | | Imports | | Exports | | Consumption | |
|---|---|---|---|---|---|---|---|---|---|---|---|---|---|---|---|---|
| | (1) | (2) | (3) | (4) | (5) | (6) | (7) | (8) | (9) | (10) | (11) | (12) | (13) | (14) | (15) | (16) |
| External risk | 0.292*** | -0.057*** | 0.530*** | 0.070 | -0.029 | -0.321*** | 0.475*** | 0.075*** | 0.249*** | -0.073*** | 0.513*** | 0.068 | -0.285* | -0.451*** | 0.448*** | 0.075*** |
| | (0.090) | (0.019) | (0.070) | (0.055) | (0.142) | (0.096) | (0.071) | (0.021) | (0.096) | (0.021) | (0.075) | (0.059) | (0.150) | (0.103) | (0.077) | (0.023) |
| Inland water | | 0.004 | | -0.191*** | | 0.116*** | | -0.029*** | | 0.003 | | -0.191*** | | 0.108*** | | -0.029*** |
| | | (0.007) | | (0.019) | | (0.034) | | (0.007) | | (0.007) | | (0.019) | | (0.034) | | (0.008) |
| Area harvested | | 0.855*** | | -0.283*** | | 1.050*** | | 0.406*** | | 0.856*** | | -0.282*** | | 1.067*** | | 0.406*** |
| | | (0.010) | | (0.028) | | (0.049) | | (0.011) | | (0.010) | | (0.028) | | (0.049) | | (0.011) |
| Population | | 0.137*** | | 1.203*** | | -0.310*** | | 0.605*** | | 0.137*** | | 1.203*** | | -0.317*** | | 0.610*** |
| | | (0.015) | | (0.044) | | (0.077) | | (0.017) | | (0.015) | | (0.044) | | (0.077) | | (0.017) |
| Yield | | 0.659*** | | 0.013 | | 1.565*** | | 0.373*** | | 0.660*** | | 0.014 | | 1.587*** | | 0.373*** |
| | | (0.019) | | (0.054) | | (0.095) | | (0.021) | | (0.019) | | (0.054) | | (0.095) | | (0.021) |
| Producer price index | | 0.031*** | | 0.123*** | | 0.212*** | | 0.079*** | | 0.021** | | 0.118*** | | 0.095** | | 0.082*** |
| | | (0.008) | | (0.022) | | (0.038) | | (0.008) | | (0.009) | | (0.025) | | (0.043) | | (0.010) |
| Disasters | | 0.034 | | 0.031 | | -0.161 | | -0.164*** | | 0.037 | | 0.030 | | -0.100 | | -0.176*** |
| | | (0.024) | | (0.067) | | (0.119) | | (0.026) | | (0.024) | | (0.069) | | (0.121) | | (0.027) |
| Droughts | | -0.109** | | -0.459*** | | 0.078 | | -0.111** | | -0.110** | | -0.467*** | | 0.083 | | -0.114** |
| | | (0.047) | | (0.135) | | (0.238) | | (0.053) | | (0.048) | | (0.137) | | (0.239) | | (0.053) |
| Floods | | 0.024 | | -0.277*** | | 0.407*** | | 0.025 | | 0.016 | | -0.282*** | | 0.301** | | 0.030 |
| | | (0.028) | | (0.079) | | (0.140) | | (0.031) | | (0.028) | | (0.081) | | (0.142) | | (0.032) |
| Year FE | N | N | N | N | N | N | N | N | Y | Y | Y | Y | Y | Y | Y | Y |
| Observation | 1,847 | 1,847 | 1,847 | 1,847 | 1,847 | 1,847 | 1,847 | 1,847 | 1,847 | 1,847 | 1,847 | 1,847 | 1,847 | 1,847 | 1,847 | 1,847 |
| $R^2$ | 0.006 | 0.958 | 0.030 | 0.451 | 0.00002 | 0.574 | 0.024 | 0.919 | 0.004 | 0.958 | 0.025 | 0.444 | 0.002 | 0.570 | 0.019 | 0.919 |

Table 7-9 The effects of internal risk on production, imports, exports and consumption

|  | Production | | Imports | | Exports | | Consumption | | Production | | Imports | | Exports | | Consumption | |
| --- | --- | --- | --- | --- | --- | --- | --- | --- | --- | --- | --- | --- | --- | --- | --- | --- |
|  | (1) | (2) | (3) | (4) | (5) | (6) | (7) | (8) | (9) | (10) | (11) | (12) | (13) | (14) | (15) | (16) |
| Internal risk | 0.135 | -0.156*** | 0.427*** | -0.090 | -0.648*** | -0.672*** | 0.296*** | -0.100*** | 0.103 | -0.184*** | 0.424*** | -0.112* | -0.849*** | -0.813*** | 0.273*** | -0.126*** |
|  | (0.089) | (0.020) | (0.070) | (0.057) | (0.140) | (0.099) | (0.071) | (0.022) | (0.095) | (0.021) | (0.074) | (0.061) | (0.146) | (0.106) | (0.076) | (0.024) |
| Inland water |  | -0.002 |  | -0.201*** |  | 0.095*** |  | -0.039*** |  | -0.004 |  | -0.202*** |  | 0.086*** |  | -0.041*** |
|  |  | (0.007) |  | (0.019) |  | (0.033) |  | (0.007) |  | (0.007) |  | (0.019) |  | (0.033) |  | (0.007) |
| Area harvested |  | 0.852*** |  | -0.287*** |  | 1.040*** |  | 0.402*** |  | 0.853*** |  | -0.285*** |  | 1.056*** |  | 0.403*** |
|  |  | (0.010) |  | (0.028) |  | (0.048) |  | (0.011) |  | (0.010) |  | (0.028) |  | (0.048) |  | (0.011) |
| Population |  | 0.161*** |  | 1.234*** |  | -0.216*** |  | 0.639*** |  | 0.165*** |  | 1.238*** |  | -0.216*** |  | 0.649*** |
|  |  | (0.015) |  | (0.044) |  | (0.077) |  | (0.017) |  | (0.016) |  | (0.045) |  | (0.078) |  | (0.018) |
| Yield |  | 0.625*** |  | -0.012 |  | 1.420*** |  | 0.346*** |  | 0.620*** |  | -0.014 |  | 1.414*** |  | 0.341*** |
|  |  | (0.019) |  | (0.055) |  | (0.097) |  | (0.022) |  | (0.019) |  | (0.056) |  | (0.097) |  | (0.022) |
| Producer price index |  | 0.031*** |  | 0.127*** |  | 0.211*** |  | 0.083*** |  | 0.020** |  | 0.117*** |  | 0.089** |  | 0.081*** |
|  |  | (0.007) |  | (0.022) |  | (0.038) |  | (0.008) |  | (0.008) |  | (0.025) |  | (0.043) |  | (0.010) |
| Disasters |  | 0.022 |  | 0.015 |  | -0.205* |  | -0.181*** |  | 0.023 |  | 0.013 |  | -0.153 |  | -0.195*** |
|  |  | (0.023) |  | (0.068) |  | (0.118) |  | (0.026) |  | (0.024) |  | (0.069) |  | (0.120) |  | (0.027) |
| Droughts |  | -0.105** |  | -0.459*** |  | 0.098 |  | -0.111** |  | -0.111** |  | -0.473*** |  | 0.087 |  | -0.121** |
|  |  | (0.047) |  | (0.135) |  | (0.236) |  | (0.053) |  | (0.047) |  | (0.137) |  | (0.236) |  | (0.053) |
| Floods |  | 0.036 |  | -0.268*** |  | 0.455*** |  | 0.034 |  | 0.030 |  | -0.275*** |  | 0.364*** |  | 0.037 |
|  |  | (0.027) |  | (0.079) |  | (0.138) |  | (0.031) |  | (0.028) |  | (0.081) |  | (0.141) |  | (0.032) |
| Year FE | N | N | N | N | N | N | N | N | Y | Y | Y | Y | Y | Y | Y | Y |
| Observations | 1,847 | 1,847 | 1,847 | 1,847 | 1,847 | 1,847 | 1,847 | 1,847 | 1,847 | 1,847 | 1,847 | 1,847 | 1,847 | 1,847 | 1,847 | 1,847 |
| $R^2$ | 0.001 | 0.959 | 0.020 | 0.452 | 0.011 | 0.581 | 0.009 | 0.919 | 0.001 | 0.959 | 0.018 | 0.445 | 0.018 | 0.579 | 0.007 | 0.919 |

Note: This table presents the results of the effects of internal risk on production, imports, exports, consumption respectively. From columns (1) to (8), all the regressions are estimated by OLS without considering year fixed effect. Columns (1), (3), (5), and (7) show regression results of internal risk impacting on production, imports, exports, consumption, respectively without concerning control variables. Columns (2), (4), (6), and (8) show regression results of internal risk impacting on production, imports, exports, consumption, respectively with concerning control variables. From columns (9) to (16), all the regressions are estimated by OLS with considering year fixed effect. Columns (9), (11), (13), and (15) show regression results of internal risk impacting on production, imports, exports, consumption, respectively without concerning control variables. Columns (10), (12), (14), and (16) show regression results of internal risk impacting on production, imports, exports, consumption, respectively with concerning control variables. The Standard Deviation is reported in round brackets. *, **, and *** are used to denote significance at the 10%, 5%, and 1% level, respectively.

Table 7-10 The effects of RSD on reserve ratio

|  | Reserve ratio | | | |
| --- | --- | --- | --- | --- |
|  | (1) | (2) | (3) | (4) |
| RSD | 1.923*** | 1.628*** | 1.830*** | 1.614*** |
|  | (0.365) | (0.314) | (0.364) | (0.317) |
| Inland water |  | 0.011 |  | 0.011 |
|  |  | (0.014) |  | (0.014) |
| Area harvested |  | 0.074*** |  | 0.072*** |
|  |  | (0.022) |  | (0.022) |
| Population |  | -0.012 |  | -0.008 |
|  |  | (0.033) |  | (0.033) |
| Yield |  | 0.877*** |  | 0.873*** |
|  |  | (0.041) |  | (0.041) |
| Producer price index |  | 0.158*** |  | 0.170*** |
|  |  | (0.016) |  | (0.019) |
| Disasters |  | -0.221*** |  | -0.237*** |
|  |  | (0.052) |  | (0.053) |
| Droughts |  | 0.160 |  | 0.137 |
|  |  | (0.103) |  | (0.104) |
| Floods |  | 0.228*** |  | 0.249*** |
|  |  | (0.060) |  | (0.062) |
| Year FE | N | N | Y | Y |
| Observations | 1,847 | 1,847 | 1,847 | 1,847 |
| $R^2$ | 0.015 | 0.382 | 0.014 | 0.370 |

Note: This table presents the results of the effects of RSD on reserve ratio. In columns (1) and (2), we show the impact of RSD on reserve ratio, without and with adding control variables using OLS regression. In columns (3) and (4), year fixed effect is taken into consideration. The Standard Deviation is reported in round brackets. *, **, and *** are used to denote significance at the 10%, 5%, and 1% level, respectively.

# A. Appendix I: data and variables

Figure A-1 World map for three types of risks

Panel A Political risk

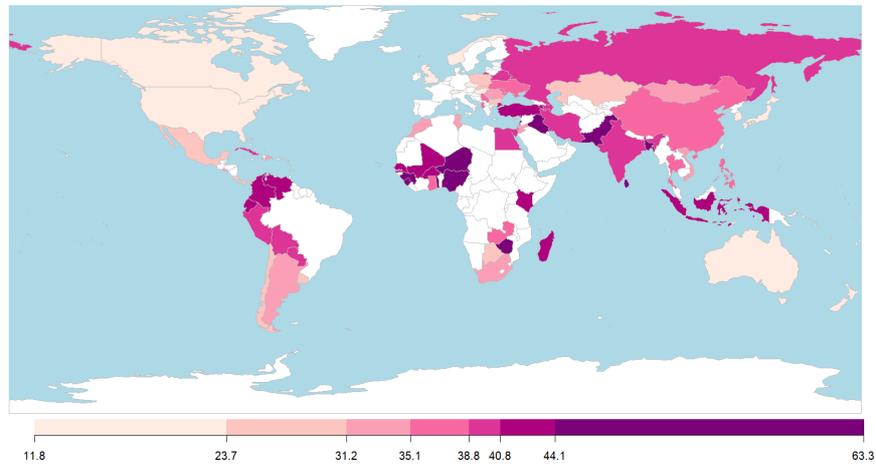

Panel B Internal risk

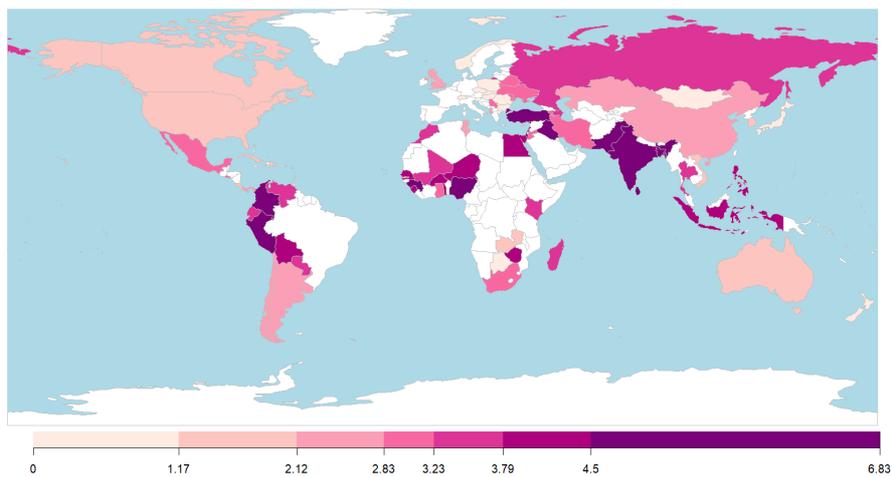

Panel C External risk

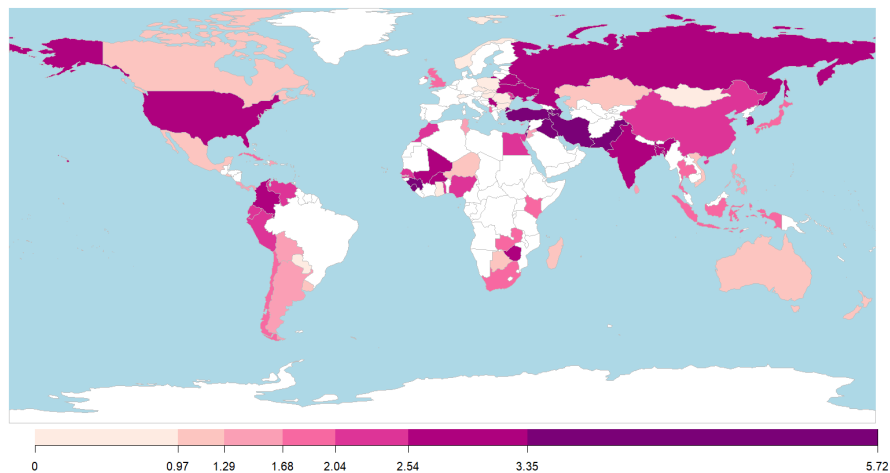

Note: we calculate the mean value for political risk, external risk, and internal risk from 1991 to 2019. Each risk increases when the color becomes darker in each Panel (Data are not available for white color parts).

Table A-1 Description of variables

| Variable | Description | Unit | Source |
|---|---|---|---|
| **Panel A:** *Dependent variables* | | | |
| Reserve ratio | $Reserve\ ratio = 100 * Beginning\ stocks/Domestic\ consumption$, where $Beginning\ stocks$ is equal to "Production + Imports – Domestic consumption - Exports". Both Beginning stocks and Domestic consumption are calculated by using five staple foods (rice, wheat and three kinds of coarse grains). | - | PSD |
| Production | Amount of production for five staple foods (rice, wheat and three kinds of coarse grains) | MT | PSD |
| Total consumption | Amount of domestic consumption plus FSI (Food, Seed and Industrial) consumption for five staple foods (rice, wheat and three kinds of coarse grains) | MT | PSD |
| Imports | Amount of imports for five staple foods (rice, wheat and three kinds of coarse grains) | MT | PSD |
| Exports | Amount of exports for five staple foods (rice, wheat and three kinds of coarse grains) | MT | PSD |
| **Panel B:** *Key variables* | | | |
| Political risk | $Political\ risk = 100 - Political\ risk_{ICRG}$, where $Political\ risk_{ICRG}$ is from ICRG. This variable is an assessment of the political stability of the countries covered by ICRG on a comparable basis, which is done by assigning score to a preset group of factors, named as political risk components. Each component has a minimum score of 0, and maximum score fixed (4,6 or 12). Specifically, there are 12 subcomponents including government stability (12), socioeconomic conditions (12), investment profile (12), internal conflict (12), external conflict (12), corruption (6), military in politics (6), religious tensions (6), law and order (6), ethnic tensions (6), democratic accountability (6), bureaucracy quality (4). Hence, political risk is scored on a scale from 0 (very low risk) to 100 (very high risk). | - | ICRG |
| External risk | $External\ risk = 12 - External\ conflict_{ICRG}$, where $External\ conflict_{ICRG}$ is from ICRG. This is an assessment of risk to the incumbent government from foreign action, ranging from non-violent external pressure (diplomatic pressures, withholding of aid, trade restrictions, territorial disputes, sanctions, *etc.*) to violent external pressure (cross-border conflicts to all-out war). The risk rating assigned is the sum of three subcomponents (including war, cross-border conflict, and foreign pressures), each with a maximum score of 4 and a minimum score of 0. Hence, external risk is scored on a scale from 0 (very low risk) to 12 (very high risk). | - | ICRG |
| Internal risk | $Internal\ risk = 12 - Internal\ conflict_{ICRG}$, where $Internal\ conflict_{ICRG}$ is from ICRG. This is an assessment of political violence in the country and its actual or potential impact on governance, and there are three subcomponents including civil war/coup threat, terrorism/political violence, and civil disorder. The risk rating assigned is the sum of three subcomponents, each with a maximum score of 4 and a minimum score of 0. Hence, internal risk is scored on a scale from 0 (very low risk) to 12 (very high risk). | - | ICRG |
| **Panel C:** *Control variables* | | | |
| Inland water | Area occupied by major rivers, lakes and reservoirs | 1000 hm$^2$ | FAOSTAT |
| Area harvested | Area harvested for five staple foods (rice, wheat and three kinds of coarse grains) | 1000 hm$^2$ | PSD |
| Population | Number of persons | Person | WBO |
| Yield | Harvested production per unit of harvested area for five staple foods (rice, wheat and three kinds of coarse grains) | MT/hm$^2$ | PSD |
| Producer price index | The agricultural producer prices measure the average annual change over time in the selling prices received by farmers (Prices at the farm-gate or at the first point of sale). We use a composite producer price index by summing the index value for five staple foods (rice, wheat and three kinds of coarse grains). | - | FAOSTAT |
| Disasters | Number of disasters | - | WDR |
| Droughts | Number of droughts | - | WDR |
| Floods | Number of floods | - | WDR |
| **Panel D:** *Grouping variables* | | | |
| Machinery | This includes 8 kinds of agricultural machinery in total. 1. Agricultural tractors, total; 2. Ploughs (e.g., reversible and non-reversible ploughs); 3. Seeders, planters and trans planters; 4. Manure spreaders and Fertilizer distributors; 5. Combine harvesters - threshers; 6. Balers (straw and fodder balers including pick-up balers); 7. Root or tuber harvesting machines; 8. Threshing machines. | - | FAOSTAT |
| GDP per capita | Gross Domestic Product per capita | USD / person | WBO |
| Import dependency ratio (IDR) | Import dependency ratio = Imports / (Domestic consumption + FSI consumption) | - | PSD, FAO |
| Self-sufficiency ratio (SSR) | Self-sufficiency ratio = Production × 100 / (Production + Imports - Exports + Ending stocks - Beginning stocks) | - | PSD |

Note: The calculation of self-sufficiency ratio is followed by Michael et al. (2015).

Table A-2 Political risk, external risk, and internal risk for sample countries

| No. | Country | Reserve ratio | Political risk | External risk | Internal risk |
|---|---|---|---|---|---|
| 1 | Paraguay | 59.37 | 39.10 | 0.93 | 3.62 |
| 2 | China | 56.60 | 35.52 | 2.41 | 2.34 |
| 3 | Australia | 52.59 | 15.21 | 1.17 | 1.41 |
| 4 | Canada | 51.36 | 13.30 | 1.13 | 1.26 |
| 5 | United States | 48.92 | 17.72 | 2.76 | 1.45 |
| 6 | Kazakhstan | 46.27 | 31.00 | 1.00 | 2.33 |
| 7 | Jordan | 45.51 | 33.07 | 1.59 | 3.17 |
| 8 | Norway | 37.07 | 12.90 | 0.90 | 0.76 |
| 9 | Uruguay | 35.21 | 28.62 | 1.07 | 2.59 |
| 10 | Thailand | 35.11 | 37.24 | 2.03 | 3.48 |
| 11 | Switzerland | 34.10 | 11.79 | 0.90 | 0.00 |
| 12 | Hungary | 33.69 | 22.75 | 0.00 | 0.00 |
| 13 | Moldova | 32.53 | 38.05 | 2.05 | 2.86 |
| 14 | Morocco | 31.13 | 32.10 | 2.07 | 3.31 |
| 15 | Iran | 28.94 | 40.59 | 4.17 | 3.03 |
| 16 | Argentina | 25.76 | 31.52 | 1.41 | 2.24 |
| 17 | Tunisia | 24.88 | 31.59 | 1.38 | 2.24 |
| 18 | Iraq | 24.54 | 63.34 | 5.72 | 6.83 |
| 19 | Republic of Korea | 24.05 | 23.66 | 3.34 | 1.62 |
| 20 | Serbia | 23.87 | 38.79 | 3.14 | 2.93 |
| 21 | Czech Republic | 23.43 | 19.83 | 0.00 | 0.17 |
| 22 | Japan | 23.18 | 18.10 | 1.69 | 1.10 |
| 23 | Philippines | 22.87 | 37.34 | 1.34 | 4.03 |
| 24 | Costa Rica | 22.37 | 26.00 | 1.66 | 1.90 |
| 25 | Ukraine | 22.10 | 37.32 | 2.55 | 2.86 |
| 26 | New Zealand | 21.30 | 13.38 | 1.00 | 0.14 |
| 27 | El Salvador | 20.81 | 34.86 | 1.28 | 2.93 |
| 28 | South Africa | 20.10 | 32.76 | 1.69 | 2.97 |
| 29 | Ecuador | 19.86 | 42.55 | 2.17 | 3.69 |
| 30 | Romania | 19.12 | 32.38 | 0.13 | 0.75 |
| 31 | Zimbabwe | 17.50 | 49.69 | 2.72 | 3.79 |
| 32 | Belarus | 17.27 | 40.23 | 2.68 | 2.91 |
| 33 | Dominican Republic | 17.14 | 34.21 | 1.52 | 2.10 |
| 34 | Egypt | 17.00 | 40.76 | 2.21 | 4.07 |
| 35 | India | 16.43 | 40.38 | 3.03 | 4.79 |
| 36 | Slovakia | 16.17 | 24.33 | 0.00 | 0.00 |
| 37 | Sri Lanka | 16.11 | 44.62 | 1.59 | 5.79 |
| 38 | Pakistan | 16.02 | 51.69 | 3.66 | 4.97 |
| 39 | Russia | 15.95 | 40.43 | 3.07 | 3.68 |
| 40 | Kenya | 15.85 | 42.17 | 1.97 | 3.76 |
| 41 | Israel | 15.65 | 35.60 | 4.64 | 4.68 |
| 42 | Azerbaijan | 15.54 | 38.81 | 4.43 | 3.29 |
| 43 | Senegal | 14.65 | 43.07 | 2.31 | 4.24 |
| 44 | Venezuela | 14.47 | 44.08 | 2.17 | 3.63 |

| 45 | United Kingdom | 14.43 | 17.50 | 2.00 | 2.75 |
|---|---|---|---|---|---|
| 46 | Peru | 13.90 | 39.55 | 2.21 | 4.79 |
| 47 | Turkey | 13.83 | 42.24 | 3.38 | 4.62 |
| 48 | Indonesia | 13.81 | 43.41 | 1.69 | 4.21 |
| 49 | Chile | 13.15 | 24.21 | 1.90 | 2.45 |
| 50 | Colombia | 12.52 | 42.24 | 2.55 | 5.86 |
| 51 | Panama | 11.76 | 31.14 | 1.59 | 2.45 |
| 52 | Poland | 11.32 | 23.75 | 0.00 | 0.63 |
| 53 | Mexico | 9.74 | 31.14 | 1.14 | 3.21 |
| 54 | Ghana | 9.61 | 35.17 | 0.62 | 3.07 |
| 55 | Bolivia | 9.43 | 39.03 | 1.38 | 3.83 |
| 56 | Armenia | 8.56 | 40.76 | 4.71 | 3.00 |
| 57 | Nigeria | 7.98 | 53.52 | 2.17 | 4.86 |
| 58 | Bulgaria | 6.98 | 29.88 | 0.38 | 0.25 |
| 59 | Mali | 6.57 | 43.79 | 2.72 | 3.76 |
| 60 | Vietnam | 5.81 | 35.10 | 1.07 | 1.76 |
| 61 | Cyprus | 5.58 | 25.63 | 1.75 | 2.00 |
| 62 | Zambia | 5.08 | 37.45 | 1.83 | 2.00 |
| 63 | Bangladesh | 5.05 | 47.90 | 2.79 | 4.62 |
| 64 | Albania | 4.45 | 35.59 | 1.90 | 2.31 |
| 65 | Niger | 4.10 | 48.97 | 1.28 | 4.03 |
| 66 | Gambia | 3.91 | 36.14 | 1.66 | 2.41 |
| 67 | Lebanon | 3.41 | 44.24 | 5.14 | 4.48 |
| 68 | Guinea | 2.76 | 50.97 | 3.90 | 4.52 |
| 69 | Botswana | 2.73 | 25.83 | 1.00 | 1.17 |
| 70 | Burkina Faso | 2.68 | 42.66 | 3.34 | 4.24 |
| 71 | Sierra Leone | 2.04 | 50.21 | 3.52 | 4.24 |
| 72 | Togo | 0.34 | 49.21 | 2.48 | 3.93 |
| 73 | Madagascar | 0.00 | 41.17 | 1.00 | 3.34 |
| 74 | Cuba | 0.00 | 40.07 | 2.03 | 2.07 |
| 75 | Mongolia | 0.00 | 31.97 | 0.21 | 1.17 |

# B. Appendix II: robustness test result for grouping variables

Table B-1 The joint effects of political risk and machinery on reserve ratio

|  | Reserve ratio | | | |
| --- | --- | --- | --- | --- |
|  | (1) | (2) | (3) | (4) |
| Political risk | -0.567** | -0.632*** | -0.635** | -0.681*** |
|  | (0.255) | (0.079) | (0.258) | (0.081) |
| Machinery | 0.014 |  | 0.003 |  |
|  | (0.094) |  | (0.095) |  |
| Machinery = 1 |  | 0.844 |  | 0.753 |
|  |  | (0.567) |  | (0.569) |
| Political risk * Machinery | 0.007 |  | 0.010 |  |
|  | (0.025) |  | (0.025) |  |
| Political risk * (Machinery = 1) |  | -0.458*** |  | -0.438*** |
|  |  | (0.167) |  | (0.167) |
| Inland water | -0.028* | -0.047*** | -0.030** | -0.050*** |
|  | (0.015) | (0.015) | (0.015) | (0.015) |
| Area harvested | 0.066*** | 0.113*** | 0.068*** | 0.110*** |
|  | (0.023) | (0.021) | (0.024) | (0.021) |
| Population | 0.044 | 0.089** | 0.051 | 0.099*** |
|  | (0.034) | (0.035) | (0.035) | (0.035) |
| Yield | 0.709*** | 0.687*** | 0.697*** | 0.671*** |
|  | (0.044) | (0.044) | (0.045) | (0.044) |
| Producer price index | 0.135*** | 0.151*** | 0.136*** | 0.161*** |
|  | (0.017) | (0.016) | (0.019) | (0.019) |
| Disasters | -0.246*** | -0.215*** | -0.257*** | -0.228*** |
|  | (0.052) | (0.051) | (0.053) | (0.052) |
| Droughts | 0.216** | 0.129 | 0.189* | 0.099 |
|  | (0.102) | (0.101) | (0.103) | (0.102) |
| Floods | 0.260*** | 0.196*** | 0.274*** | 0.217*** |
|  | (0.061) | (0.060) | (0.062) | (0.061) |
| Year FE | N | N | Y | Y |
| Observations | 1,847 | 1,847 | 1,847 | 1,847 |
| $R^2$ | 0.398 | 0.414 | 0.388 | 0.406 |

Table B-2 The joint effects of political risk and GDP per capita on reserve ratio

|  | Reserve ratio | | | |
| --- | --- | --- | --- | --- |
|  | (1) | (2) | (3) | (4) |
| Political risk | -0.981** | -0.779*** | -1.071*** | -0.883*** |
|  | (0.385) | (0.114) | (0.400) | (0.119) |
| GDP per capita | -0.062 |  | -0.093 |  |
|  | (0.144) |  | (0.148) |  |
| GDP per capita = 1 |  | -1.913*** |  | -2.224*** |
|  |  | (0.577) |  | (0.586) |
| Political risk * GDP per capita | 0.120*** |  | 0.129*** |  |
|  | (0.039) |  | (0.040) |  |
| Political risk * (GDP per capita = 1) |  | 0.596*** |  | 0.680*** |
|  |  | (0.174) |  | (0.176) |
| Inland water |  | -0.057*** | -0.027* | -0.058*** | -0.029* |

|  | (0.014) | (0.015) | (0.014) | (0.015) |
|---|---|---|---|---|
| Area harvested | 0.133*** | 0.115*** | 0.131*** | 0.115*** |
|  | (0.020) | (0.021) | (0.020) | (0.021) |
| Population | 0.063* | 0.043 | 0.071** | 0.051 |
|  | (0.032) | (0.035) | (0.033) | (0.035) |
| Yield | 0.489*** | 0.724*** | 0.482*** | 0.713*** |
|  | (0.045) | (0.045) | (0.045) | (0.045) |
| Producer price index | 0.109*** | 0.145*** | 0.125*** | 0.148*** |
|  | (0.016) | (0.016) | (0.018) | (0.019) |
| Disasters | -0.220*** | -0.235*** | -0.241*** | -0.245*** |
|  | (0.049) | (0.052) | (0.050) | (0.053) |
| Droughts | 0.295*** | 0.181* | 0.271*** | 0.148 |
|  | (0.097) | (0.102) | (0.098) | (0.103) |
| Floods | 0.257*** | 0.242*** | 0.279*** | 0.257*** |
|  | (0.057) | (0.060) | (0.059) | (0.062) |
| Year FE | N | N | Y | Y |
| Observations | 1,847 | 1,847 | 1,847 | 1,847 |
| $R^2$ | 0.456 | 0.398 | 0.448 | 0.389 |

Table B-3 The joint effects of political risk and IDR on reserve ratio

|  | Reserve ratio | | | |
|---|---|---|---|---|
|  | (1) | (2) | (3) | (4) |
| Political risk | -0.385*** | -0.646*** | -0.419*** | -0.696*** |
|  | (0.123) | (0.098) | (0.126) | (0.100) |
| IDR | 4.641** |  | 4.566** |  |
|  | (1.941) |  | (1.957) |  |
| IDR = 1 |  | -0.363 |  | -0.427 |
|  |  | (0.457) |  | (0.459) |
| Political risk * IDR | -0.602 |  | -0.589 |  |
|  | (0.520) |  | (0.524) |  |
| Political risk * (IDR = 1) |  | 0.183 |  | 0.199 |
|  |  | (0.124) |  | (0.125) |
| Inland water | 0.015 | -0.017 | 0.013 | -0.020 |
|  | (0.015) | (0.015) | (0.016) | (0.015) |
| Area harvested | 0.306*** | 0.170*** | 0.302*** | 0.167*** |
|  | (0.030) | (0.027) | (0.030) | (0.027) |
| Population | -0.153*** | 0.004 | -0.144*** | 0.013 |
|  | (0.041) | (0.038) | (0.042) | (0.038) |
| Yield | 0.853*** | 0.737*** | 0.839*** | 0.723*** |
|  | (0.046) | (0.044) | (0.046) | (0.045) |
| Producer price index | 0.117*** | 0.139*** | 0.127*** | 0.140*** |
|  | (0.016) | (0.016) | (0.018) | (0.019) |
| Disasters | -0.258*** | -0.253*** | -0.271*** | -0.264*** |
|  | (0.050) | (0.051) | (0.052) | (0.052) |
| Droughts | 0.247** | 0.225** | 0.222** | 0.198* |
|  | (0.100) | (0.102) | (0.101) | (0.103) |
| Floods | 0.294*** | 0.258*** | 0.315*** | 0.275*** |
|  | (0.059) | (0.060) | (0.060) | (0.061) |
| Year FE | N | N | Y | Y |

|             | Observations | 1,847 | 1,847 | 1,847 | 1,847 |
|-------------|--------------|-------|-------|-------|-------|
|             | R²           | 0.423 | 0.402 | 0.413 | 0.393 |

Table B-4 The joint effects of political risk and SSR on reserve ratio

|  | Reserve ratio | | | |
|---|---|---|---|---|
|  | (1) | (2) | (3) | (4) |
| Political risk | -0.545*** | -0.500*** | -0.586*** | -0.536*** |
|  | (0.076) | (0.080) | (0.078) | (0.082) |
| SSR | 0.001 |  | 0.001 |  |
|  | (0.001) |  | (0.001) |  |
| SSR = 1 |  | 0.982* |  | 1.093** |
|  |  | (0.513) |  | (0.516) |
| Political risk * SSR | -0.0003 |  | -0.0003 |  |
|  | (0.0003) |  | (0.0003) |  |
| Political risk * (SSR = 1) |  | -0.284** |  | -0.312** |
|  |  | (0.142) |  | (0.142) |
| Inland water | -0.027* | -0.028* | -0.029** | -0.031** |
|  | (0.015) | (0.015) | (0.015) | (0.015) |
| Area harvested | 0.110*** | 0.104*** | 0.109*** | 0.101*** |
|  | (0.022) | (0.024) | (0.022) | (0.024) |
| Population | 0.048 | 0.057 | 0.056 | 0.066* |
|  | (0.035) | (0.036) | (0.035) | (0.036) |
| Yield | 0.724*** | 0.721*** | 0.711*** | 0.707*** |
|  | (0.045) | (0.044) | (0.045) | (0.045) |
| Producer price index | 0.143*** | 0.145*** | 0.147*** | 0.149*** |
|  | (0.016) | (0.016) | (0.019) | (0.019) |
| Disasters | -0.248*** | -0.247*** | -0.260*** | -0.260*** |
|  | (0.052) | (0.052) | (0.053) | (0.053) |
| Droughts | 0.198* | 0.194* | 0.171* | 0.165 |
|  | (0.103) | (0.103) | (0.104) | (0.104) |
| Floods | 0.257*** | 0.255*** | 0.276*** | 0.274*** |
|  | (0.060) | (0.060) | (0.061) | (0.061) |
| Year FE | N | N | Y | Y |
| Observations | 1,847 | 1,847 | 1,847 | 1,847 |
| R² | 0.394 | 0.395 | 0.384 | 0.386 |

# C. Appendix III: endogeneity test result

Table C-1 The effects of political risk on reserve ratio (lag 3 years)

|  | Reserve ratio | | | |
|---|---|---|---|---|
|  | (1) | (2) | (3) | (4) |
| Political risk | -1.080*** | -0.606*** | -1.115*** | -0.637*** |
|  | (0.075) | (0.076) | (0.075) | (0.078) |
| Inland water |  | -0.034** |  | -0.036** |
|  |  | (0.015) |  | (0.015) |
| Area harvested |  | 0.107*** |  | 0.107*** |
|  |  | (0.022) |  | (0.022) |
| Population |  | 0.050 |  | 0.056 |
|  |  | (0.036) |  | (0.036) |
| Yield |  | 0.702*** |  | 0.695*** |
|  |  | (0.046) |  | (0.046) |
| Producer price index |  | 0.142*** |  | 0.140*** |
|  |  | (0.017) |  | (0.019) |
| Disasters |  | -0.215*** |  | -0.222*** |
|  |  | (0.053) |  | (0.054) |
| Droughts |  | 0.220** |  | 0.210** |
|  |  | (0.106) |  | (0.107) |
| Floods |  | 0.238*** |  | 0.246*** |
|  |  | (0.062) |  | (0.064) |
| Year FE | N | N | Y | Y |
| Observations | 1,697 | 1,697 | 1,697 | 1,697 |
| $R^2$ | 0.109 | 0.395 | 0.116 | 0.384 |

Table C-2 The effect of external risk and internal risk on reserve ratio (lag 3 years)

| | Reserve ratio | | | | | | | | | | | |
|---|---|---|---|---|---|---|---|---|---|---|---|---|
| | (1) | (2) | (3) | (4) | (5) | (6) | (7) | (8) | (9) | (10) | (11) | (12) |
| External risk | -0.026 | | 0.111* | -0.109** | | -0.076 | -0.128** | | 0.009 | -0.131*** | | -0.095* |
| | (0.052) | | (0.058) | (0.043) | | (0.048) | (0.055) | | (0.060) | (0.046) | | (0.049) |
| Internal risk | | -0.248*** | -0.299*** | | -0.116** | -0.082* | | -0.331*** | -0.334*** | | -0.140*** | -0.105** |
| | | (0.052) | (0.059) | | (0.045) | (0.050) | | (0.054) | (0.059) | | (0.048) | (0.052) |
| Inland water | | | | -0.006 | -0.007 | -0.009 | | | | -0.008 | -0.009 | -0.012 |
| | | | | (0.015) | (0.015) | (0.015) | | | | (0.015) | (0.015) | (0.015) |
| Area harvested | | | | 0.107*** | 0.106*** | 0.105*** | | | | 0.107*** | 0.105*** | 0.104*** |
| | | | | (0.022) | (0.022) | (0.022) | | | | (0.022) | (0.023) | (0.022) |
| Population | | | | -0.030 | -0.022 | -0.017 | | | | -0.024 | -0.015 | -0.007 |
| | | | | (0.035) | (0.036) | (0.036) | | | | (0.036) | (0.036) | (0.036) |
| Yield | | | | 0.841*** | 0.819*** | 0.823*** | | | | 0.843*** | 0.815*** | 0.820*** |
| | | | | (0.043) | (0.044) | (0.044) | | | | (0.043) | (0.045) | (0.045) |
| Producer price index | | | | 0.166*** | 0.163*** | 0.165*** | | | | 0.166*** | 0.165*** | 0.165*** |
| | | | | (0.017) | (0.017) | (0.017) | | | | (0.019) | (0.019) | (0.019) |
| Disasters | | | | -0.148*** | -0.150*** | -0.153*** | | | | -0.158*** | -0.162*** | -0.166*** |
| | | | | (0.054) | (0.054) | (0.054) | | | | (0.055) | (0.055) | (0.055) |
| Droughts | | | | 0.151 | 0.154 | 0.153 | | | | 0.141 | 0.143 | 0.140 |
| | | | | (0.107) | (0.107) | (0.107) | | | | (0.108) | (0.108) | (0.108) |
| Floods | | | | 0.200*** | 0.208*** | 0.207*** | | | | 0.206*** | 0.218*** | 0.214*** |
| | | | | (0.063) | (0.063) | (0.063) | | | | (0.065) | (0.065) | (0.065) |
| Year FE | N | N | N | N | N | N | Y | Y | Y | Y | Y | Y |
| Observations | 1,697 | 1,697 | 1,697 | 1,697 | 1,697 | 1,697 | 1,697 | 1,697 | 1,697 | 1,697 | 1,697 | 1,697 |
| $R^2$ | 0.0001 | 0.013 | 0.015 | 0.374 | 0.374 | 0.375 | 0.003 | 0.022 | 0.022 | 0.362 | 0.363 | 0.364 |

In the first stage, we use the Probit regression model to estimate binary political risk dummy, which equals 1 if political risk is larger than mean value, and 0 otherwise. We add the same control variables as in previous regression, control year fixed effect and generate Inverse Mills ratio (IMR). See Equation (1) for this Probit regression equation. In the second stage, we use the Logit regression model to estimate binary reserve ratio dummy, which equals 1 if reserve ratio is larger than mean value, and 0 otherwise. We add IMR as an independent variable, add same control variables, and also control year fixed effect. Note that in second stage, we allow for one year lag in all independent variables. See Equation (2) for this Logit regression equation. We also replace political risk with external/internal risk and do the same test. The results can be seen in Table C-3, Table C-4 and Table C-5.

$$Probit(binary-Political\ risk)_{i,t} = \sum_{k=1}^{8} \beta_k * \log(1+X^k)_{i,t} + c_i + year_t + \varepsilon_{i,t} \qquad (1)$$

$$Logit(binary-Reserve\ ratio)_{i,t} = \log(1+Political\ risk)_{i,t-1} + IMR_{i,t-1} + \sum_{k=1}^{8} \beta_k * \log(1+X^k)_{i,t-1} + c_i + year_{t-1} + \varepsilon_{i,t-1} \qquad (2)$$

Table C-3 Heckman two stage regression for political risk

|  | Binary – Political risk | Binary – Reserve ratio |
|---|---|---|
|  | *Probit* | *Logistic* |
|  | *stage 1* | *stage 2* |
| Political risk |  | -2.134*** |
|  |  | (0.247) |
| IMR |  | -0.600 |
|  |  | (0.964) |
| Inland water | -0.079*** | -0.036 |
|  | (0.020) | (0.055) |
| Area harvested | -0.019 | 0.098* |
|  | (0.030) | (0.057) |
| Population | 0.485*** | 0.022 |
|  | (0.047) | (0.275) |
| Yield | -0.685*** | 1.405*** |
|  | (0.059) | (0.386) |
| Producer price index | -0.123*** | 0.199** |
|  | (0.025) | (0.082) |
| Disasters | -0.368*** | -0.204 |
|  | (0.073) | (0.242) |
| Droughts | 0.222 | 0.035 |
|  | (0.146) | (0.294) |
| Floods | 0.114 | 0.328* |

|  | (0.086) | (0.173) |
|---|---|---|
| Year FE | Y | Y |
| Observations | 1,922 | 1,847 |

Table C-4 Heckman two stage regression for external risk

|  | Binary – External risk | Binary – Reserve ratio |
|---|---|---|
|  | *Probit* | *Logistic* |
|  | *stage 1* | *stage 2* |
| External risk |  | -0.095 |
|  |  | (0.108) |
| IMR |  | 1.725 |
|  |  | (1.190) |
| Inland water | -0.160*** | -0.148 |
|  | (0.021) | (0.110) |
| Area harvested | -0.091*** | 0.005 |
|  | (0.030) | (0.078) |
| Population | 0.456*** | 0.364 |
|  | (0.047) | (0.301) |
| Yield | 0.101* | 1.598*** |
|  | (0.058) | (0.135) |
| Producer price index | -0.031 | 0.172*** |
|  | (0.024) | (0.052) |
| Disasters | -0.163** | -0.333** |
|  | (0.075) | (0.169) |
| Droughts | 0.015 | 0.026 |
|  | (0.149) | (0.266) |
| Floods | 0.055 | 0.379** |
|  | (0.088) | (0.160) |
| Year FE | Y | Y |
| Observations | 1,922 | 1,847 |

Table C-5 Heckman two stage regression for internal risk

|  | Binary – Internal risk | Binary – Reserve ratio |
|---|---|---|
|  | *Probit* | *Logistic* |
|  | *stage 1* | *stage 2* |
| Internal risk |  | -0.490*** |
|  |  | (0.125) |
| IMR |  | 1.482 |
|  |  | (1.335) |
| Inland water | -0.139*** | -0.133 |
|  | (0.020) | (0.112) |
| Area harvested | -0.037 | 0.031 |
|  | (0.029) | (0.064) |
| Population | 0.480*** | 0.453 |
| | (0.046) | (0.385) |
| Yield | -0.393*** | 1.062*** |
| | (0.057) | (0.324) |
| Producer price index | 0.025 | 0.233*** |
| | (0.024) | (0.049) |
| Disasters | -0.300*** | -0.476* |
| | (0.072) | (0.270) |
| Droughts | -0.054 | -0.048 |
| | (0.142) | (0.269) |
| Floods | 0.125 | 0.465** |
| | (0.085) | (0.183) |
| Year FE | Y | Y |
| Observations | 1,922 | 1,847 |